\documentclass{article}

    \PassOptionsToPackage{numbers, compress}{natbib}

\usepackage[preprint]{neurips_2025}

\usepackage[utf8]{inputenc} 
\usepackage[T1]{fontenc}    
\usepackage{hyperref}       
\usepackage{url}            
\usepackage{booktabs}       
\usepackage{amsfonts}       
\usepackage{nicefrac}       
\usepackage{microtype}      
\usepackage{xcolor}         

\usepackage{algpseudocode}
\usepackage{algorithm}
\usepackage{thmtools}
\usepackage{amsmath}
\usepackage{amssymb}
\usepackage{amsthm}
\usepackage{graphicx}
\usepackage{subcaption}

\title{Converging to Stability in Two-Sided Bandits: The Case of Unknown Preferences on Both Sides of a Matching Market}

\author{
    Gaurab Pokharel\\ 
    Virginia Tech\\
    Alexandria, VA\\
    \texttt{gaurab@vt.edu}
    \And 
    Sanmay Das\\
    Virginia Tech\\ 
    Alexandria, VA\\
    \texttt{sanmay@vt.edu}
}

\newtheorem{lemma}{Lemma}[section]
\newtheorem{sublemma}{Lemma}[lemma]  

\newtheorem{definition}{Definition}

\newcommand{\UCB}{\text{UCB}}
\newcommand{\LCB}{\text{LCB}}
\newcommand{\argmax}{\mathop{\mathrm{arg\,max}}\limits}
\newcommand{\DeltaMin}{\Delta_{\min}}

\newcommand{\smallcall}[2]{\Call{\tiny #1}{#2}}

\makeatletter
\newcommand{\resetlinenumber}{\setcounter{ALG@line}{0}}
\makeatother

\newcommand{\optimismfxn}{%
  f(x)=
    \begin{cases}
      \dfrac{1-e^{-\kappa x}}{1-e^{-\kappa/2}}, & 0\le x\le 0.5,\\[1ex]
      1, & x>0.5
    \end{cases}%
}

\begin{document}

\maketitle

\begin{abstract}
We study the problem of repeated two-sided matching with uncertain preferences (two-sided bandits), and no explicit communication between agents. Recent work has developed algorithms that converge to stable matchings when one side (the proposers or agents) must learn their preferences, but the preferences of the other side (the proposees or arms) are common knowledge, and the matching mechanism uses simultaneous proposals at each round. We develop new algorithms that provably converge to stable matchings for two more challenging settings: one where the arm preferences are no longer common knowledge, and a second, more general one where the arms are also uncertain about their preferences. In our algorithms, agents start with optimistic beliefs about arms' preferences and update these preferences over time. The key insight is in how to combine these beliefs about arm preferences with beliefs about the value of matching with an arm conditional on one's proposal being accepted when choosing whom to propose to. 
\end{abstract}

\section{Introduction} \label{sec_introduction}

The classic literature on two-sided matching \cite[e.g.]{GS, Roth_Xing_1997, Haeringer_Wooders_2011}, encompassing applications including long- and short-term labor markets, dating and marriage, school choice, and more, has typically focused on situations where agents are aware of their own preferences. The problem of learning preferences while participating in a repeated matching market first started receiving attention in the AI literature in the work of \citet{two_sided_matching}, and the general idea of two-sided matching under unknown preferences has since been studied in economics and operations research as well ~\cite{Lee_Schwarz_2009,johari2022experimental}. This area of research has received renewed attention in the last few years, along with novel theoretical insights into convergence properties of upper-confidence-bound (UCB) style algorithms \cite{caucb,Kong_Yin_Li_2022,communicative}.

The two-sided matching problem involves agents on two sides of a market who have preferences for each other but cannot communicate explicitly. The goal is to create a matching process that ensures \emph{stability}, where no pairs of agents would rather be matched with each other over their current match. The existence of such matchings was famously demonstrated constructively in the Gale-Shapley algorithm \cite{GS} which is structured around one side of the market \emph{proposing} and the other side choosing whether to accept proposals. This theory has been applied to various markets, such as matching medical students to residencies \cite{roth1999redesign} and students to schools \cite{abdulkadirouglu2005boston}, with the assumption that agents know their own preferences. Considerable interest has also been shown in the AI community regarding two-sided matching in the presence of various constraints, such as diversity constraints \cite{aziz2022matching,benabbou2018diversity}. A version of the problem, called \emph{two-sided bandits}, was introduced in \citet{two_sided_matching}, where agents engage in repeated matching without prior knowledge of preferences, and the impact of the matching mechanism on \emph{convergence} to stability was studied. Recent work has also focused on computing the probability that a matching is stable or finding the matching with the highest probability of being stable under different models of preference uncertainty \cite{Aziz_and_co}.

The exploration-exploitation dilemma that characterizes bandit problems is further complicated in two-sided bandits by uncertainty on not just the values of the ``arms'' of the bandit, but also uncertainty on whether the arms will accept or reject your attempt to pull them (your proposal). This additional uncertainty arises from the multi-agent nature of the problem; instead of there being just one player, there are multiple players competing for the arms, and the arms also have preferences associated with the players. 
A popular solution choice for this type of problem involves the explore-then-commit style of algorithms \cite{pagare2023twosided, zhang2024decentralized}. This flavor of solution includes a first phase of ``learning'' preferences, and a second phase using what has been learned in the prior one to commit to a choice. This involves spending some dedicated time trying to figure out the preferences of the agents. In this paper we are more interested in approaches where learning happens simultaneously with the matching process. One such algorithm, CA-UCB \cite{caucb}, assumes that at each round all the agents on the receiving side (henceforth \emph{arms}) get to view all their proposals before deciding which one to accept. The algorithm provably converges to stability under the assumptions that (1) all arms have complete knowledge of their preferences, and (2) these preferences are common knowledge, so the proposers (henceforth \emph{players}) also know them. The algorithm is a variation of the well-known family of UCB algorithms. Complementary to this, \citet{Hosseini_Roy_Zhang_2024} take a \emph{stability-centric} view on the problem: using the \emph{arm-proposing} version of the Gale–Shapley matching procedure, they show higher likelihood of stable outcomes and provide Probably Approximately Correct (PAC) bounds on the sample complexity of reaching stability. They do this by employing an action-elimination strategy to rule out clearly inferior arms when an agent faces multiple options, concentrating on pairs most likely to appear in a stable matching.

\subsubsection*{Contributions:} We study the more general case where neither the players nor the arms start off with knowledge of their own preferences. This problem is significantly more complex than  the case with one-sided uncertainty and common knowledge of the other side's preferences. 
We approach the problem in two steps. 

First, we relax the common knowledge assumption and assume only that arms are aware of their own preferences. We design an algorithm (Optimistic CA-UCB or OCA-UCB) that provably converges to a stable matching in this setting, using an extension of the concept of ``plausible sets'' (arms that an player might be able to ``win'' in the next time step) from the CA-UCB algorithm \cite{caucb}. Our work differs from other recent papers that also do not assume common knowledge; those rely on either explicit communication \cite{communicative} or implicit communication through carefully orchestrated ``collisions'' \cite{zhang2024decentralized}. 


Second, we introduce the Probabilistic Conflict Avoiding - Simultaneous Choice Algorithm (PCA-SCA) for the case with uncertainty on both sides. PCA-SCA leverages an additional optimistic estimate for the probability of winning each contentious proposal. We prove convergence to stability using UCB-style reward estimates. The framework can also accommodate other methods such as Thompson Sampling. Experimentally, we demonstrate convergence to stability under both UCB and Thompson-based reward estimates and evaluate how the convergence speed scales with the number of agents and the degree of preference heterogeneity.

\section{Setting} \label{sec_problem}

There are $N$ proposers (henceforth players) and $K$ proposees (henceforth arms). To ensure that each player can be matched, we assume that $N \leq K$ \cite{centralized,caucb,basu21,sankararaman21}. The set of players is denoted by $\left\{p_i\right\}_{i=1}^N$ and the set of arms by $\left\{a_k\right\}_{k=1}^K$. For each $p_i$, there is a distinct (unknown) mean reward $\mu_i^k$ associated with each arm $a_k$ with unit variance. If $\mu_i^k > \mu_i^j$, we say that $p_i$ prefers $a_k$ over $a_j$ denoted by $a_k \succ_{p_i} a_j$. Similarly, the arms also have distinct mean rewards associated with each player. The arms may or may not have this information available to them a priori depending on the setting in consideration. 

At each time-step $t \in \{1, \cdots, T\}$, player $p_i$ attempts to pull an arm $a_k$. Let $A_i(t)$ and $\bar{A}_i(t)$ denote, respectively, the arm that player $p_i$ attempted to pull, and the arm that player $p_i$ successfully pulled at time $t$. If multiple players attempt to pull the same arm, a conflict will arise. If the arms are aware of their preferences, the player that is the most preferred by the arm will be picked in the event of a conflict (if the arms are not fully aware of their preferences, then the chosen player will depend on the arm's decision-making algorithm). At the end of each time-step, if a player $p_i$ is successfully matched with an arm i.e. $\bar{A}_i(t) =  m_i(t)$, it will receive a stochastic reward $X_i^{m_i(t)} \sim \mathcal{N}(\mu_i^{m_i(t)}, 1)$.  Players that fail to successfully pull an arm will receive a reward of 0 and $\bar{A}_i(t) = \emptyset$. The final matching is made public to all the agents at the end of each time step.

We use the classic notion of a stable matching: a bipartite matching from the group of players to the group of arms such that no player and arm prefer each other over their current matching i.e $\nexists (p_i, m_i(t))$ such that $p_i$ prefers an arm $a_j \neq m_i(t)$ and $a_j$ also prefers $p_i$ over its current match. There may be more than one stable match. Our primary focus in this work is on convergence to stability. Another quantity we study is maximum Player Pessimal Regret. We use the Gale-Shapley algorithm with arms as proposers in order to calculate the player-pessimal stable match \cite{GS}. Denote the reward received by player $p_i$ in a player pessimal match by $X_i$. Then, we define maximum player-regret at each time-step as $R(t) = \underset{i \in N}{\max} \left[X_i - X_i^{m(t)} \right]$ which is the maximum difference (over players) between the reward a player received in its current match and the reward it would have received in a stable matching that player-pessimal.

In this work we consider matching markets with varying levels of information availability. As is standard in models of learning in two-sided matching, we assume no explicit communication between players on the proposing side. Throughout the rest of the paper we assume that the players do not have access to their own preferences and need to learn them over time. As mentioned above, we consider several scenarios in terms of arm's preferences. For convenience, we name those as follows: 

\begin{itemize}
    \item Scenario 0: Players have access to arm preferences. Arms have access to arm preferences. We call this the \textbf{Arm Preferences Common Knowledge (APCK)} model.
    \item Scenario 1: Players do not have access to arm preferences. Arms have access to arm preferences. We call this the \textbf{Arm Preferences Known but Private (APKP)} model.
    \item Scenario 2:  Players do not have access to arm preferences. Arms do not have access to arm preferences. We call this the \textbf{Arm Preferences Unknown (APU)} model.
\end{itemize}

\noindent The APCK model has been studied and a decentralized solution approach is described in the Conflict Avoiding Upper Confidence Bound (CA-UCB) algorithm \cite{caucb}. Our main contribution is to detail decentralized approaches to solutions for Scenarios 1 and 2. We begin with a quick recap of CA-UCB in the APCK model.
\paragraph{APCK Model and CA-UCB} \label{section:problem}

The APCK model (Scenario 0) was studied in a recent paper \cite{centralized}. Subsequently, the CA-UCB algorithm was proposed as a decentralized solution \cite{caucb}. CA-UCB finds a stable match in the APCK model by avoiding conflicts using the notion of a ``plausible set.'' The plausible set for a player $p_i$ at time-step $t$ is defined as $S_i(t) := \{a_k : p_i \succ_{a_k} p_j, \text{ where } \bar{A}_j(t-1) = a_k\}$ i.e. the set of all arms $a_k$ such that at time-step $(t-1)$ another player $p_j \neq p_i$ successfully pulled the arm, but $p_i$ knows that $a_k$ prefers $p_i$ over $p_j$. This set also includes the arms the player successfully pulled (if any) and all the arms that were unmatched at the previous time-step. This makes it so that $p_i$ only attempts arms where there is a possibility of it successfully pulling the arm, avoiding conflicts. Then, $p_i$ attempts to pull the arm with the highest UCB value among $S_i(t)$ . The UCB value tracked by $p_i$ for an arm $a_k$ is calculated as $\UCB_i^k = \hat{\mu}_i^k + \sqrt{\frac{3\ln(t)}{2n_i^k(t)}}$, 
where $\hat{\mu}_i^k$ is the empirical mean reward tracked by $p_i$ for $a_k$ up to time $t$, and $n_i^k(t)$ is the total number of samples that is used to calculate that empirical mean. 
\section{APKP Model and OCA-UCB} \label{sec_APKP_model}

CA‐UCB requires both players and arms to know arms’ preferences in order to form $S_i(t)$.  When players lack this information, they must learn their rankings in arms' preferences and and infer plausible sets.  Recent decentralized methods address this:  Etesami and Srikant \cite{Etesami_Srikant_2024} maintain and update proposal‐arm probability distributions via accept/reject feedback, yielding stable matchings with logarithmic regret in \emph{hierarchical} markets; Shah et al. \cite{Shah_Ferguson_Marden_2024} employ ``mood'' states (content, discontent, watchful) to guide proposals based on utility feedback, converging to the proposer‐optimal stable match without centralized coordination. We introduce OCA‐UCB (Optimistic CA‐UCB), a simpler approach that exploits arms’ self‐knowledge and needs at most one ``collision'' per player to learn arm preferences.

Following the general structure of CA‐UCB, each player initializes $\mathrm{UCB}_i^k=\infty$.  Additionally, for each arm $a_k$, player $p_i$ maintains two sets, 
$O^{k_h}_{(i)}=\{p_j\mid p_j\succ_{a_k}p_i\}$ and $O^{k_l}_{(i)}=\{p_j\mid p_i\succ_{a_k}p_j\}$ i.e. the set of players $a_k$ ranks above and below $p_i$ in $a_k$'s preference order.  
These start optimistically as $O^{k_h}_{(i)}=\emptyset$ and $O^{k_l}_{(i)}=\{p_j\}_{j\neq i}$, i.e.\ each player assumes it is every arm’s top choice. Notice that to update plausible sets, $p_i$ requires only knowing, when $\bar A_j(t-1)=a_k$, which of $p_i$ or $p_j$ is preferred by $a_k$. Because the initial estimates are optimistic, all arms are plausible and conflicts become inevitable.  If at time $t$ players $p_i$ and $p_j$ conflict over $a_k$ and $p_j$ wins, then $O^{k_h}_{(i)}\gets O^{k_h}_{(i)}\cup\{p_j\}$ and $O^{k_l}_{(i)}\gets O^{k_l}_{(i)}\setminus\{p_j\}$. Thereafter $p_i$ excludes $p_j$ when forming $S_i(t)$. Apart from this update, the mechanism is identical to CA-UCB. Eventually, each player’s sets recover the arms’ true orderings. We prove using Theorem \ref{thm:simple_theorem} (stated below) that, using belief updates as mentioned above preserves CA-UCB's guarantee of convergence. The proof is detailed in Appendix \ref{appendix:oca_ucb_proof}.

\begin{restatable}{theorem}{SimpleTheorem}
    \label{thm:simple_theorem}
    Under the belief update scheme described above, OCA-UCB shares CA-UCB's guarantee of convergence to stability. 
\end{restatable}

\section{APU Model and PCA-SCA} \label{sec:APU_Model}

The previous section studied the APKP model, where players lacked direct knowledge of arms' preferences but leveraged deterministic feedback—obtained from arms that perfectly knew their preferences—to reconstruct a plausible set akin to CA-UCB. In contrast, under the APU model, arms themselves do not know their preferences, leading to noisy conflict feedback and invalidating the previous approach. Due to this unreliable conflict feedback, we propose an alternative approach enabling agents to select optimal arms while avoiding conflicts. We introduce a selection rule based on maximizing the \emph{expected} reward combined with estimated probabilities of winning conflicts. Each player maintains estimates of its conflict-win probabilities against other players, choosing the arm that maximizes the product of these probabilities with reward estimates.

Simultaneously, each arm maintains empirical reward estimates and confidence intervals for all players. These intervals serve two purposes: when intervals are disjoint, the arm deterministically identifies the highest-reward player; when intervals overlap, the arm resolves conflicts by selecting among overlapping players uniformly at random. This approach ensures convergence. The full algorithm (Algorithm~\ref{alg:pca_daa}) can be found in the Appendix.

\noindent The algorithm uses two parameters, $\lambda\in[0,1)$ and $\kappa\ge 1$. $\lambda$ injects randomized delays (line~\ref{line:bernoulli_sample} of \textproc{PCA-SCA}) by sampling $D_i(t)\sim\mathrm{Bernoulli}(\lambda)\,$ to reduce the likelihood of conflicts  \cite{caucb,Kong_Yin_Li_2022}. If $D_i(t)=0$, player $i$ selects the arm with the highest expected reward; otherwise, it repeats its previous choice.  $\kappa$ tunes the level of \emph{optimism} in the player’s conflict–win probability estimates (line~\ref{line:optimism} of \textproc{GetBestArm}) -- the specifics of which are detailed below.  All reward estimates are initialized to $\infty$ and all win probabilities to 1. First let us go over how the arms pick the players at each round.

\paragraph{Arms' Choice of Players: } At each time step, players submit proposals to the arms according to a specified protocol (detailed next). Each arm, upon receiving one or more proposals, must select exactly one proposing player. Because the arms initially do not know their preferences, each arm maintains beliefs about the payoff of each player and confidence intervals for those beliefs.

Each arm processes the incoming proposals as follows. If at least one player’s reward estimate is still infinite (i.e., uninitialized), the arm selects uniformly at random from those players (Line \ref{line:infinite_players} of \textproc{ResolvePullRequests}). Otherwise, the arm identifies the player with the highest upper confidence bound and checks whether other players’ confidence intervals overlap with this best bound (Line \ref{line:check_overlap}). If multiple intervals overlap, the arm selects uniformly at random from among those candidates (Line \ref{line:best_players}); if only one player remains with the highest bound, it selects that player.

\paragraph{Players' Choice of Arms:} 
Each player initializes all arm‐reward estimates to $\infty$ and all empirical conflict‐win probabilities $\hat P_{ij}^k=1$, updating each $\hat P_{ij}^k$ to the fraction of conflicts won.  At each time step, player $p_i$ selects the arm that maximizes its expected reward $\arg\max_{k} \Bigl[\hat{P}_{ij}^k \cdot \mathcal{R}_i^k\Bigr]$, where $j$ is the index of the player who pulled arm $k$ in the previous step, and $\mathcal{R}_i^k$ is $p_i$’s current reward estimate for arm $a_k$. When there is overlap, arms choose uniformly at random, so in expectation $\hat{P}_{ij}^k\to 0.5$. Once intervals separate, selection is deterministic and $\hat{P}_{ij}^k$ converges to its true value. To prevent premature arm abandonment from transiently low estimates, we introduce the \emph{optimism function}:





\begin{equation}
  \label{eqn:optimism_fxn}
  \optimismfxn
\end{equation}

\noindent Here $\kappa\ge1$ controls the level of optimism. As $\kappa\to\infty$, $f(\hat P_{ij}^k)\to1$ for all $\hat P_{ij}^k$, whereas $\kappa=1$ yields a nearly linear yet optimistic taper.  Once feedback is deterministic and $\hat P_{ij}^k\to p_{ij}^k$, any $\hat P_{ij}^k\ge0.5$ is boosted to 1 when $p_{ij}^k=1$, while for $p_{ij}^k=0$, $f(\hat P_{ij}^k)$ decays slowly at a rate determined by $\kappa$.  Thus, $\kappa$ directly controls the aggressiveness of optimism (see Figure~\ref{fig:optimism} in Appendix \ref{appendix:optimism_viz}).



At each round, every player selects the arm that maximizes the product of its reward estimate and its optimistic win‐probability (line \ref{line:circ} of \textproc{GetBestArm}).  Then each arm chooses the player with the highest expected reward (mirroring the simultaneous‐choice rule of \cite{two_sided_matching}).  Successful pulls generate mutual rewards and update both arms’ reward estimates and players’ win‐probability beliefs. 

Finally, note that the structure of our algorithm accommodates multiple methods for updating beliefs about expected rewards and conflict win probabilities. Here, we test UCB \cite{caucb} and Thompson Sampling \cite{Kong_Yin_Li_2022} for belief updates in simulation, in addition to providing a proof of convergence to stability when UCB is used.




\subsection{Using UCB in PCA-SCA (PCA-UCB)}

We integrate the classic Upper Confidence Bound (UCB) \cite{original_ucb} method into PCA-SCA (see Algorithm~\ref{alg:pca_daa}) and refer to it as PCA-UCB.  Each player maintains

\begin{equation}
    \tag{2}
    \text{UCB}_i^k(t) = 
        \begin{cases}
            \infty & \qquad \text{ if } n_i^k(t) = 0\\
            \hat{\mu}_i^k + \sqrt{\frac{3 \ln (t)}{2 n_i^k (t-1)}} & \qquad \text{ otherwise}
        \end{cases}
        \label{eq:ucb_reward_player}
\end{equation}
\noindent where $n_i^k(t)$ is the number of times player $p_i$ has pulled arm $a_k$ at time-step $t$, and $\hat{\mu}_i^k$ is the empirical mean tracked by player $p_i$ for arm $a_k$. Let $c_i^k(t) = \sqrt{\frac{3 \ln (t)}{2 n_i^k (t-1)}}$, then each arm tracks for every player both a lower and upper confidence bounds given by $\LCB_i^k(t) = \hat{\mu}_i^k(t) - c_i^k(t)$ and $\UCB_i^k(t) = \hat{\mu}_i^k(t) + c_i^k(t)$. Similar to the players, the LCB and the UCB for the arms is initialized to $-\infty$ and $\infty$ respectively. The other quantity that the players keep track of is the probability of conflict wins: 

\begin{equation}
    \tag{3}
    \label{equ:probability_entry}
    \hat{P}_{ij}^k(t) =
        \begin{cases}
                1 & \qquad \text{ if } n_{ij}^k = 0\\
                \frac{w_{ij}^k(t)}{N_{c_{ij}}^k(t)} & \qquad \text{ otherwise}
        \end{cases}
\end{equation}

\noindent Here, $\hat P_{ij}^k(t)$ is $p_i$’s estimated probability of defeating $p_j$ on arm $a_k$, where $w_{ij}^k(t)$ and $N_{c_{ij}}^k(t)$ are the counts of wins and total conflicts.  After each round—once rewards and matchings are revealed—the \textproc{Update-Win-Probability} function (Algorithm~\ref{alg:pca_daa}) increments the estimate on a win and decrements it on a loss.  

To state our convergence theorems, we introduce the following notation and definitions. We assume, without loss of generality, that players are ordered so that $\mu_i^k > \mu_j^k$ when $p_i \succ_{a_k} p_j$. We say that the confidence intervals for arm $a_k$ \textbf{are \emph{disjoint}} if, for every pair $\{p_i, p_j\}$ with $\mu_i^k > \mu_j^k$, $\LCB_i^k(t) > \UCB_j^k(t)$. Conversely, if for some $\{p_i, p_j\}$ with $\mu_i^k > \mu_j^k$ we have $ \LCB_i^k(t) \leq \UCB_j^k(t),$ then the confidence intervals for those two players are said to be \textbf{overlapping}. Next, the deterministic feedback event is defined as: 

\begin{definition}[Deterministic Feedback Event $\alpha(t)$]
    \label{def:alpha_t}
    For any time $t$, define
    \begin{equation*}
        \alpha(t) = \Bigl\{ \forall\, k\in[K],\, \forall\, i,j \text{ with } \mu_i^k > \mu_j^k:\; \LCB_i^k(t) > \UCB_j^k(t) \Bigr\}.
    \end{equation*}
    Thus, $\alpha(t)$ is the event that, for every arm, the players’ confidence intervals are disjoint at time $t$. When $\alpha(t)$ holds, the arm’s feedback is deterministic.
\end{definition}

 CA-UCB assumes known preferences, and OCA-UCB learns them deterministically—treating arms in $S_i(t)$ as having win probability of 1. In APU, plausibility generalizes to any arm with positive win probability. Therefore, we define \emph{inconsistency} in the APU model as follows.

\begin{definition}[Inconsistent Triplet]
\label{dfn:inconsistent_triplet}
    Let $P_{ij}^k$ denote the true probability of player $p_i$ winning against player $p_j$ in a conflict for arm $a_k$, and let $\hat{P}_{ij}^k$ be the corresponding empirical probability. The triplet $(p_i, p_j, a_k)$ is said to be \emph{inconsistent} if
        \begin{equation*}
            \Bigl[ \Bigl\{ \Bigl(P_{ij}^k =1 \Bigr) \wedge \Bigl( \hat{P}_{ij}^k \leq 1 - \epsilon \Bigr) \Bigr\} \vee  \Bigl\{ \Bigl(P_{ij}^k = 0 \Bigr) \wedge \Bigl( \hat{P}_{ij}^k \geq \epsilon \Bigr) \Bigr\} \Bigr],
        \end{equation*}
\end{definition}

\begin{definition}[Inconsequential Triplet]
    \label{dfn:inconsequantial_triplet}
    An inconsistent triplet is considered \emph{inconsequential} if there exists an alternative arm that offers a better reward prospect. Formally, consider two triplets $(p_i, p_j, a_k)$ and $(p_i, p_j', a_k')$. If, at time $t$, the triplet $(p_i, p_j, a_k)$ is inconsistent but $\UCB_i^{k'} > \UCB_i^{k}$ and the triplet $(p_i, p_j', a_k')$ is consistent (i.e., arm $a_{k'}$ is a better prospect for $p_i$), then player $p_i$ will select $a_{k'}$. In this case, the inconsistency in $(p_i, p_j, a_k)$ is rendered inconsequential.
\end{definition}
\noindent A triplet is considered \emph{consequential} if it is not inconsequential. Finally, for any pair of players $p_i$, $p_j$, and $a_k$, define:
\begin{equation*}
    \begin{array}{rcl}
        \Delta_{ij}^k & = & \mu_i^k - \mu_j^k \quad \text{for } \mu_i^k > \mu_j^k, \\
        \DeltaMin & =& \underset{i \neq j, \forall k}{\min} \lvert\mu_i^k - \mu_j^k \rvert, \\
        n_{ij}^k(t) &=& \min\{n_i^k(t), n_j^k(t)\}.
    \end{array}
\end{equation*}

Note that, without loss of generality, we assume $\DeltaMin = 1$.  With the key definitions in place, we can now state our three main convergence theorems. Their proofs are deferred to the Appendix.

\begin{restatable}{theorem}{TheoremOne}[Finite-Time Learning of Arms' Preferences] 
    \label{thm:1}
    In an instance of the APU model with $K$ arms and $N$ players running PCA-UCB, at time $t$, with probability at least $1 - \delta(t)$, for every arm, the confidence intervals that the arm tracks for each player are disjoint. Here $\delta(t) = K \cdot \frac{N^2}{2} \cdot \exp\left( -\frac{ n(t) \left( \Delta_{\text{min}} - 2 c(t) \right)^2 }{ 4 } \right)$
    \noindent where $n(t) = \underset{\substack{ i \in N \\ k \in K}}{\min} \left\{n_i^k(t) \right\}$, $\DeltaMin = \underset{\substack{i,j \in N\\ k \in K}}{\min} \left\{\Delta_{ij}^k \right\}$, and $c(t) = \sqrt{\frac{3 \ln(t)}{2n(t)}}$.
\end{restatable}

 Our first result shows that arms rapidly learn their true preference rankings, so that after a finite number of samples, conflicts between players send deterministic feedback.

\begin{restatable}{theorem}{TheoremTwo}[Player Win Estimate Convergence]
    \label{thm:2}
    Assume that event $\alpha$ has occurred by time $T$. Then, at time $T + t$, with probability at least $1 - \tau (T)$  for all \emph{consequential} conflicts $(p_i, p_j, a_k)$ (defined above): 
    \begin{equation*}
        \lvert \hat{P}_{ij}^k (T+t) - p_{ij}^k \rvert \leq \frac{T}{T+t}
    \end{equation*}    
    \noindent Here $\tau(t) = \frac{KN^2}{2} \exp\left(-2\, N_c(T)\, \left( \left( \Delta_{\text{min}} - 2 c(T)  \right) \sqrt{\frac{n(T)}{8N_c(T)}} \right)^2\right)$, 
    where $N_c(t)$ is the minimum number of conflicts across all triplets. 
\end{restatable}

 Theorem~\ref{thm:2} guarantees that once feedback becomes noise‐free, players’ empirical win‐probabilities converge to their true values at a $O(T/(T+t))$ rate.

\begin{restatable}{theorem}{TheoremThree}[Identical Execution to CA-UCB] 
    \label{thm:3}
    After time $T + t$, the execution of PCA-UCB is identical to that of CA-UCB. 
\end{restatable}

 Together, these theorems imply that PCA-UCB not only learns preferences and refines win‐estimates efficiently, but ultimately replicates the behavior of the idealized CA-UCB algorithm in the long run.

\subsection{Using Thompson Sampling in PCA-SCA (PCA-TS)}


Thompson Sampling \cite{original_thompson} (TS) is an alternative MAB method that has resurged in recent literature \cite{agrawal2012analysis,chapelle2011empirical,Kong_Yin_Li_2022}. In the PCA-SCA algorithm, TS tracks reward beliefs for both players and arms; we call this variant PCA-TS. Like UCB, both players and arms record total rewards, pulls, and conflict wins per player. However, TS updates beliefs differently: players compute conflict win estimates as in Equation~\ref{equ:probability_entry} and then apply Equation~\ref{eqn:optimism_fxn} for an optimistic estimate. As for the rewards, we assume the variance ($\sigma^2 = 1$) in rewards of the arms and the players are known. Then, we update the mean and precision ($\tau = \frac{1}{\sigma^2}$) as:
\begin{equation}
    \tag{4}
    \mu_{\text{new}}, \tau_{\text{new}} =\frac{\tau_0 \mu_0 + \tau \sum_{i=1}^n x_i}{\tau_0 + n \tau} , \tau_0 + n \tau 
    \label{eq:thupdate}
\end{equation}

\noindent where $\{\mu_{\text{new}}, \tau_{\text{new}}\}$ are the updated mean and precision, $\{\mu_0, \tau_0\}$ the previous values, $\tau=1$ the known precision, and $\sum x_i$ the total reward. We bound the confidence intervals by $\mathcal{R}_i^L(t)$ and $\mathcal{R}_i^U(t)$ \footnote{
In our experiments below, we use $\mathcal{R}_i^L(t) = \mu_0 - \frac{1}{\tau_0},\mathcal{R}_i^U(t) = \mu_0 + \frac{1}{\tau_0}$ where $\frac{1}{\tau_0}$ represents the standard deviation in belief.}(see Lines \ref{line:upper} and \ref{line:lower} in \textproc{ResolvePullRequests} of Algorithm~\ref{alg:pca_daa}). When an agent estimates a reward for a player or arm, it samples from $\mathcal{N}(\mu_0, 1/\tau_0)$ using its tracked $\mu_0$ and $\tau_0$. Simulation results for PCA-TS under varying market sizes and preference heterogeneity are provided in Section~\ref{sec:results}.

\section{Simulation Results} \label{sec:results}

We simulate\footnote{See Appendix \ref{appendix:specs} for the specifications of the machine the experiments were run on} two scenarios: (1) uniformly random preferences with market sizes $N=K\in\{5,10,15,20\}$; and (2) varying preference heterogeneity at $N=K=10$.  Maximum rewards range from $1$ (least preferred) to $K+1$ (most preferred) with constant gaps $\Delta=1$.  Player heterogeneity is controlled as in \cite{Ashlagi_Kanoria_Leshno_2017,caucb} via $\beta\in\{0,10,100,1000\}$—larger $\beta$ yields less heterogeneity, with $\beta\to\infty$ implying identical preferences.  Specifically, we sample $x_k \sim \mathrm{Uniform}(0,1)$ and $\epsilon_{k,i} \sim \mathrm{Logistic}(0,1)$, set $\overline{\mu}_i^k=\beta x_k+\epsilon_{k,i}$, and define $\mu_i^k=\#\{j:\overline{\mu}_i^j\le\overline{\mu}_i^k\}$ to maintain $\Delta=1$.  Hyperparameters are $\lambda=0.9$ and $\kappa=10$ in PCA-SCA.\footnote{%
There are alternative methods for controlling heterogeneity in player preferences, for example notions like uncoordinated/coordinated markets \cite{roglin_and_colleagues}. The method we use corresponds to uncoordinated markets when $\beta$ is low. However, 
coordinated markets are different, with correlation defined on the edges of the matching graph rather than the vertices. We find that the convergence time in these edge-correlated cases is actually comparable to when preferences are uniformly random, compared with node-correlation, where convergence time increases (see Appendix).
}

To objectively quantify the quality and rate of convergence, we introduce a notion of a ``convergence proxy'': It measures, in sliding windows of size $\mathcal{X}$, the percentage of time-steps where market stability was greater than $\theta$. It effectively measures the tendency of a market to converge within the next $\mathcal{X}$ time-steps. We calculate the ``convergence proxy'' line for PCA-UCB vs PCA-TS for random preferences and varied market sizes. This allows us to objectively visualize the convergence behavior of the two approaches and draw conclusions about their behaviors.

We run simulations in each of the aforementioned scenarios 100 times and monitor the market state. More specifically, we take a snap-shot of the maximum player regret and market stability every 10 time-steps for each experiment and average the results over all the runs. To evaluate the trend in the quantities we study, Loess smoothing was run on the data points to yield the graphs presented in the following sections.

\subsection{OCA-UCB in the APKP model} \label{subsec:APKP_model_results}

Recall that, in this scenario, players must learn both their own and the arms’ preference orderings. We compare OCA-UCB in the APKP model to CA-UCB in the APCK model, running 6000 steps for random preferences and 3000 for varied heterogeneity. Figure \ref{fig:APKP_results} (top) shows that OCA-UCB converges to stability—with only modest slowdowns in larger markets—closely matching CA-UCB when preferences are publicly known. The bottom row illustrates that varying player heterogeneity has negligible impact on convergence. Together with our theory, these results imply that private arm preferences pose little extra difficulty for achieving stability when arms themselves know their own orderings.


\begin{figure}[ht]
    \begin{center}  
        \includegraphics[width=0.9\textwidth]{APCK_vs_APKP.pdf}
    \end{center}
    \caption{\begin{small}
    Left: APCK model with CA-UCB; Right: APKP model with OCA-UCB. Results averaged over 100 runs. Top row: varying market size with uniformly random preferences; bottom row: varying player preference heterogeneity at $N=K=10$. OCA-UCB converges to stability under dual-sided uncertainty, though more slowly due to increased complexity.
    \end{small}}
    \label{fig:APKP_results}
\end{figure}

\subsection{PCA-UCB and PCA-TS in the APU Model} \label{subsec:APU_Model_results}
We now turn to the considerably more challenging setting where both the players and the arms do not have access to any preference information. Recall that in our approach, where at each time-step the players attempt the arm with the  \emph{highest expected reward} using the \emph{optimism function}. As the arms become more confident about their reward estimates for players, the probability estimates that the players keep track of will be more representative of true probabilities. Given the structure of Algorithm \ref{alg:pca_daa} (PCA-SCA), we evaluate two methods to keep track of agents' beliefs: UCB (PCA-UCB), and Thompson Sampling (PCA-TS). In this section, we detail the results of our experiments when using these approaches.

\begin{figure}[ht]
  \centering
  \begin{subfigure}[b]{0.45\textwidth}
    \centering
    \includegraphics[clip,width=\textwidth]{figs/PCA_UCB.pdf}
    \caption{UCB}
    \label{fig:APU_UCB}
  \end{subfigure}
  \hfill
  \begin{subfigure}[b]{0.45\textwidth}
    \centering
    \includegraphics[clip,width=\textwidth]{figs/PCA_TS.pdf}
    \caption{Thompson Sampling}
    \label{fig:APU_th}
  \end{subfigure}
  \caption{\begin{small}Results of APU model experiments using (a) UCB and (b) Thompson Sampling to track beliefs. The top row shows uniformly random preferences; the bottom row shows varied player preference heterogeneity ($N=K=10$). Both markets converge to stability in expectation and player regret vanishes; Thompson Sampling achieves faster, smoother convergence.\end{small}}
  \label{fig:APU_combined}
  \vspace{-0.5cm}
\end{figure}

\paragraph{Using PCA-UCB}
Figure \ref{fig:APU_UCB} summarizes the results of PCA-UCB in the APU model. 20,000 time steps were run for random preferences and 10,000 for varied player preferences. The first row shows results for random preferences, while the second row shows results for varied player preferences.


Figures~\ref{fig:APKP_results} and~\ref{fig:APU_UCB} show that the APU model with PCA-UCB stabilizes slower than the APKP model with OCA-UCB. This is because both players and arms need to learn beliefs from scratch. Nevertheless, with random market preferences, both models exhibit similar trends—stabilization probabilities approach 1 and player regret tends to 0—demonstrating that PCA-UCB can achieve decentralized stable matching despite unknown preferences. The second row in Figure~\ref{fig:APU_UCB} shows that varying player preference heterogeneity (measured by $\beta$) does not affect convergence, paralleling CA-UCB's behavior.

\paragraph{Using PCA-TS} 
Results for when PCA-TS was used in the APU model are shown in Figure \ref{fig:APU_th}. Like Figure \ref{fig:APU_UCB} (the UCB results), the first row in Figure \ref{fig:APU_th} is when the agent preferences are sampled uniformly at random, and the second row is where the players' preference heterogeneity is varied. Results from PCA-UCB (Figure \ref{fig:APU_UCB})  and PCA-TS (Figure \ref{fig:APU_th}) show similarities in market stability and player regret for all market sizes with uniformly random preferences, with larger markets stabilizing slower. We can also see that PCA-TS converges faster and more smoothly than PCA-UCB. We will discuss this particular property in further detail in the next subsection. The results demonstrate that Algorithm \ref{alg:pca_daa} can find a stable match in the APU model using both UCB and Thompson Sampling.


\paragraph{Convergence Comparison: UCB vs. Thompson Sampling} 

\begin{figure}[ht]
    \centering
    \includegraphics[clip,width=.45\columnwidth]{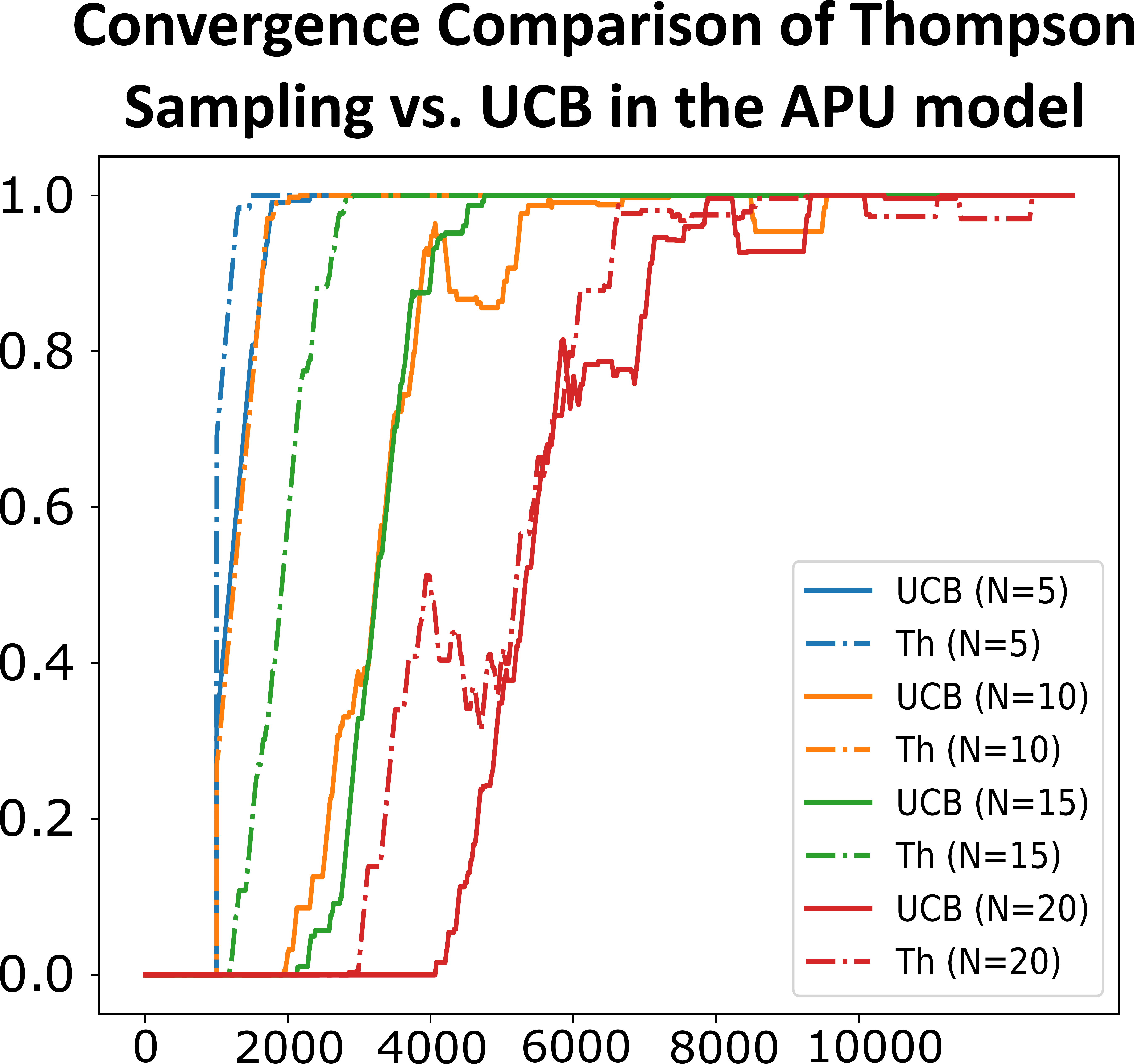}
    \caption{\begin{small}Convergence comparison of UCB (solid) and Thompson Sampling (dotted) in the APU model with varying market sizes and uniformly random preferences. Thompson Sampling converges slightly faster and remains stable, while UCB exhibits intermittent dips before ultimately converging.\end{small}}
    \label{fig:convergence_proxy}
\end{figure}

When looking at Figures~\ref{fig:APU_UCB} and~\ref{fig:APU_th}, we observe that the Thompson Sampling approach for tracking beliefs outperforms UCB in terms of both convergence speed and smoothness. To quantify this behavior, we employ the \emph{convergence proxy} defined earlier, now comparing the stability graphs of PCA-UCB and PCA-TS. Setting $\mathcal{X} = 1000$ and $\theta = 90$, we generate the corresponding lines, with the results summarized in Figure~\ref{fig:convergence_proxy}. Each color corresponds to a different market size, with the solid line representing PCA-UCB and the dotted line representing PCA-TS. Evidently, the lines for Thompson Sampling reach 1.0 (i.e., 100\% of the time-steps exhibit stability $> 90\%$) faster and remain at that level.

UCB's convergence proxy approaches 1.0 more slowly and often dips before recovering, because exploration disrupts players' belief orderings and leads to unstable matches—especially early on. In contrast, PCA-TS only shows this behavior for \(N=20\) and converges more smoothly overall. This suggests that Thompson Sampling is a better method for tracking agent beliefs in the APU model.

\section{Conclusion and Future Work} \label{sec_conclusion} 

In this paper, we describe algorithms that allow agents to learn their preferences in repeated two-sided matching. Our algorithms are the first to provably converge without explicit communication between agents (the proposers) when arm preferences are not common knowledge (the APKP model) and when arms themselves do not know their preferences (the APU model). We also demonstrate, through simulation, that using Thompson Sampling converges to stability more quickly and smoothly compared to the UCB approach in the APKP model across a range of settings.


\bibliographystyle{ACM-Reference-Format} 
\bibliography{references}


\begin{thebibliography}{28}


\ifx \showCODEN    \undefined \def \showCODEN     #1{\unskip}     \fi
\ifx \showDOI      \undefined \def \showDOI       #1{#1}\fi
\ifx \showISBNx    \undefined \def \showISBNx     #1{\unskip}     \fi
\ifx \showISBNxiii \undefined \def \showISBNxiii  #1{\unskip}     \fi
\ifx \showISSN     \undefined \def \showISSN      #1{\unskip}     \fi
\ifx \showLCCN     \undefined \def \showLCCN      #1{\unskip}     \fi
\ifx \shownote     \undefined \def \shownote      #1{#1}          \fi
\ifx \showarticletitle \undefined \def \showarticletitle #1{#1}   \fi
\ifx \showURL      \undefined \def \showURL       {\relax}        \fi
\providecommand\bibfield[2]{#2}
\providecommand\bibinfo[2]{#2}
\providecommand\natexlab[1]{#1}
\providecommand\showeprint[2][]{arXiv:#2}

\bibitem[\protect\citeauthoryear{Abdulkadiro{\u{g}}lu, Pathak, Roth, and S{\"o}nmez}{Abdulkadiro{\u{g}}lu et~al\mbox{.}}{2005}]%
        {abdulkadirouglu2005boston}
\bibfield{author}{\bibinfo{person}{Atila Abdulkadiro{\u{g}}lu}, \bibinfo{person}{Parag~A Pathak}, \bibinfo{person}{Alvin~E Roth}, {and} \bibinfo{person}{Tayfun S{\"o}nmez}.} \bibinfo{year}{2005}\natexlab{}.
\newblock \showarticletitle{The {Boston} public school match}.
\newblock \bibinfo{journal}{\emph{American Economic Review}} \bibinfo{volume}{95}, \bibinfo{number}{2} (\bibinfo{year}{2005}), \bibinfo{pages}{368--371}.
\newblock


\bibitem[\protect\citeauthoryear{Ackermann, Goldberg, Mirrokni, R\"{o}glin, and V\"{o}cking}{Ackermann et~al\mbox{.}}{2008}]%
        {roglin_and_colleagues}
\bibfield{author}{\bibinfo{person}{Heiner Ackermann}, \bibinfo{person}{Paul~W. Goldberg}, \bibinfo{person}{Vahab~S. Mirrokni}, \bibinfo{person}{Heiko R\"{o}glin}, {and} \bibinfo{person}{Berthold V\"{o}cking}.} \bibinfo{year}{2008}\natexlab{}.
\newblock \showarticletitle{Uncoordinated Two-Sided Matching Markets}. In \bibinfo{booktitle}{\emph{Proceedings of the 9th ACM Conference on Electronic Commerce}} (Chicago, Il, USA) \emph{(\bibinfo{series}{EC '08})}. \bibinfo{publisher}{Association for Computing Machinery}, \bibinfo{address}{New York, NY, USA}, \bibinfo{pages}{256–263}.
\newblock
\showISBNx{9781605581699}
\urldef\tempurl%
\url{https://doi.org/10.1145/1386790.1386831}
\showDOI{\tempurl}


\bibitem[\protect\citeauthoryear{Agrawal and Goyal}{Agrawal and Goyal}{2012}]%
        {agrawal2012analysis}
\bibfield{author}{\bibinfo{person}{Shipra Agrawal} {and} \bibinfo{person}{Navin Goyal}.} \bibinfo{year}{2012}\natexlab{}.
\newblock \showarticletitle{Analysis of {Thompson Sampling} for the multi-armed bandit problem}. In \bibinfo{booktitle}{\emph{Conference on Learning Theory}}. JMLR Workshop and Conference Proceedings.
\newblock


\bibitem[\protect\citeauthoryear{Ashlagi, Kanoria, and Leshno}{Ashlagi et~al\mbox{.}}{2017}]%
        {Ashlagi_Kanoria_Leshno_2017}
\bibfield{author}{\bibinfo{person}{Itai Ashlagi}, \bibinfo{person}{Yash Kanoria}, {and} \bibinfo{person}{Jacob~D. Leshno}.} \bibinfo{year}{2017}\natexlab{}.
\newblock \showarticletitle{Unbalanced Random Matching Markets: The Stark Effect of Competition}.
\newblock \bibinfo{journal}{\emph{Journal of Political Economy}} \bibinfo{volume}{125}, \bibinfo{number}{1} (\bibinfo{date}{Feb} \bibinfo{year}{2017}), \bibinfo{pages}{69–98}.
\newblock
\showISSN{0022-3808, 1537-534X}
\urldef\tempurl%
\url{https://doi.org/10.1086/689869}
\showDOI{\tempurl}


\bibitem[\protect\citeauthoryear{Auer, Cesa-Bianchi, and Fischer}{Auer et~al\mbox{.}}{2002}]%
        {original_ucb}
\bibfield{author}{\bibinfo{person}{Peter Auer}, \bibinfo{person}{Nicolò Cesa-Bianchi}, {and} \bibinfo{person}{Paul Fischer}.} \bibinfo{year}{2002}\natexlab{}.
\newblock \showarticletitle{Finite-time Analysis of the Multiarmed Bandit Problem}.
\newblock \bibinfo{journal}{\emph{Machine Learning}} \bibinfo{volume}{47}, \bibinfo{number}{2} (\bibinfo{date}{May} \bibinfo{year}{2002}), \bibinfo{pages}{235–256}.
\newblock
\showISSN{1573-0565}
\urldef\tempurl%
\url{https://doi.org/10.1023/A:1013689704352}
\showDOI{\tempurl}


\bibitem[\protect\citeauthoryear{Aziz, Bir{\'o}, and Yokoo}{Aziz et~al\mbox{.}}{2022}]%
        {aziz2022matching}
\bibfield{author}{\bibinfo{person}{Haris Aziz}, \bibinfo{person}{P{\'e}ter Bir{\'o}}, {and} \bibinfo{person}{Makoto Yokoo}.} \bibinfo{year}{2022}\natexlab{}.
\newblock \showarticletitle{Matching market design with constraints}. In \bibinfo{booktitle}{\emph{Proceedings of the AAAI Conference on Artificial Intelligence}}, Vol.~\bibinfo{volume}{36}. \bibinfo{pages}{12308--12316}.
\newblock


\bibitem[\protect\citeauthoryear{Aziz, Biró, Gaspers, de~Haan, Mattei, and Rastegari}{Aziz et~al\mbox{.}}{2020}]%
        {Aziz_and_co}
\bibfield{author}{\bibinfo{person}{Haris Aziz}, \bibinfo{person}{Péter Biró}, \bibinfo{person}{Serge Gaspers}, \bibinfo{person}{Ronald de Haan}, \bibinfo{person}{Nicholas Mattei}, {and} \bibinfo{person}{Baharak Rastegari}.} \bibinfo{year}{2020}\natexlab{}.
\newblock \showarticletitle{Stable Matching with Uncertain Linear Preferences}.
\newblock \bibinfo{journal}{\emph{Algorithmica}} \bibinfo{volume}{82}, \bibinfo{number}{5} (\bibinfo{date}{May} \bibinfo{year}{2020}), \bibinfo{pages}{1410–1433}.
\newblock
\showISSN{0178-4617, 1432-0541}
\urldef\tempurl%
\url{https://doi.org/10.1007/s00453-019-00650-0}
\showDOI{\tempurl}


\bibitem[\protect\citeauthoryear{Basu, Sankararaman, and Sankararaman}{Basu et~al\mbox{.}}{2021}]%
        {basu21}
\bibfield{author}{\bibinfo{person}{Soumya Basu}, \bibinfo{person}{Karthik~Abinav Sankararaman}, {and} \bibinfo{person}{Abishek Sankararaman}.} \bibinfo{year}{2021}\natexlab{}.
\newblock \showarticletitle{Beyond $log^2(T)$ regret for decentralized bandits in matching markets}. In \bibinfo{booktitle}{\emph{Proceedings of the 38th International Conference on Machine Learning}} \emph{(\bibinfo{series}{Proceedings of Machine Learning Research}, Vol.~\bibinfo{volume}{139})}, \bibfield{editor}{\bibinfo{person}{Marina Meila} {and} \bibinfo{person}{Tong Zhang}} (Eds.). \bibinfo{publisher}{PMLR}, \bibinfo{pages}{705--715}.
\newblock
\urldef\tempurl%
\url{https://proceedings.mlr.press/v139/basu21a.html}
\showURL{%
\tempurl}


\bibitem[\protect\citeauthoryear{Benabbou, Chakraborty, Ho, Sliwinski, and Zick}{Benabbou et~al\mbox{.}}{2018}]%
        {benabbou2018diversity}
\bibfield{author}{\bibinfo{person}{Nawal Benabbou}, \bibinfo{person}{Mithun Chakraborty}, \bibinfo{person}{Xuan-Vinh Ho}, \bibinfo{person}{Jakub Sliwinski}, {and} \bibinfo{person}{Yair Zick}.} \bibinfo{year}{2018}\natexlab{}.
\newblock \showarticletitle{Diversity Constraints in Public Housing Allocation}. In \bibinfo{booktitle}{\emph{Proceedings of the 17th International Conference on Autonomous Agents and MultiAgent Systems}}. \bibinfo{pages}{973--981}.
\newblock


\bibitem[\protect\citeauthoryear{Chapelle and Li}{Chapelle and Li}{2011}]%
        {chapelle2011empirical}
\bibfield{author}{\bibinfo{person}{Olivier Chapelle} {and} \bibinfo{person}{Lihong Li}.} \bibinfo{year}{2011}\natexlab{}.
\newblock \showarticletitle{An empirical evaluation of {Thompson Sampling}}.
\newblock \bibinfo{journal}{\emph{Advances in Neural Information Processing Systems}}  \bibinfo{volume}{24} (\bibinfo{year}{2011}).
\newblock


\bibitem[\protect\citeauthoryear{Das and Kamenica}{Das and Kamenica}{2005}]%
        {two_sided_matching}
\bibfield{author}{\bibinfo{person}{Sanmay Das} {and} \bibinfo{person}{Emir Kamenica}.} \bibinfo{year}{2005}\natexlab{}.
\newblock \showarticletitle{Two-Sided Bandits and the Dating Market}. In \bibinfo{booktitle}{\emph{Proceedings of the 19th International Joint Conference on Artificial Intelligence}} (Edinburgh, Scotland) \emph{(\bibinfo{series}{IJCAI'05})}. \bibinfo{publisher}{Morgan Kaufmann Publishers Inc.}, \bibinfo{address}{San Francisco, CA, USA}, \bibinfo{pages}{947–952}.
\newblock


\bibitem[\protect\citeauthoryear{Etesami and Srikant}{Etesami and Srikant}{2024}]%
        {Etesami_Srikant_2024}
\bibfield{author}{\bibinfo{person}{S.~Rasoul Etesami} {and} \bibinfo{person}{R. Srikant}.} \bibinfo{year}{2024}\natexlab{}.
\newblock \showarticletitle{Decentralized and Uncoordinated Learning of Stable Matchings: A Game-Theoretic Approach}.
\newblock  \bibinfo{number}{arXiv:2407.21294} (\bibinfo{date}{Aug.} \bibinfo{year}{2024}).
\newblock
\urldef\tempurl%
\url{https://doi.org/10.48550/arXiv.2407.21294}
\showDOI{\tempurl}
\newblock
\shownote{arXiv:2407.21294 [cs, eess].}


\bibitem[\protect\citeauthoryear{Gale and Shapley}{Gale and Shapley}{1962}]%
        {GS}
\bibfield{author}{\bibinfo{person}{D. Gale} {and} \bibinfo{person}{L.~S. Shapley}.} \bibinfo{year}{1962}\natexlab{}.
\newblock \showarticletitle{College Admissions and the Stability of Marriage}.
\newblock \bibinfo{journal}{\emph{The American Mathematical Monthly}} \bibinfo{volume}{69}, \bibinfo{number}{1} (\bibinfo{date}{Jan} \bibinfo{year}{1962}), \bibinfo{pages}{9}.
\newblock
\showISSN{00029890}
\urldef\tempurl%
\url{https://doi.org/10.2307/2312726}
\showDOI{\tempurl}


\bibitem[\protect\citeauthoryear{Haeringer and Wooders}{Haeringer and Wooders}{2011}]%
        {Haeringer_Wooders_2011}
\bibfield{author}{\bibinfo{person}{Guillaume Haeringer} {and} \bibinfo{person}{Myrna Wooders}.} \bibinfo{year}{2011}\natexlab{}.
\newblock \showarticletitle{Decentralized job matching}.
\newblock \bibinfo{journal}{\emph{International Journal of Game Theory}} \bibinfo{volume}{40}, \bibinfo{number}{1} (\bibinfo{date}{Feb} \bibinfo{year}{2011}), \bibinfo{pages}{1–28}.
\newblock
\showISSN{0020-7276, 1432-1270}
\urldef\tempurl%
\url{https://doi.org/10.1007/s00182-009-0218-x}
\showDOI{\tempurl}


\bibitem[\protect\citeauthoryear{Hosseini, Roy, and Zhang}{Hosseini et~al\mbox{.}}{2024}]%
        {Hosseini_Roy_Zhang_2024}
\bibfield{author}{\bibinfo{person}{Hadi Hosseini}, \bibinfo{person}{Sanjukta Roy}, {and} \bibinfo{person}{Duohan Zhang}.} \bibinfo{year}{2024}\natexlab{}.
\newblock \showarticletitle{Putting Gale \& Shapley to Work: Guaranteeing Stability Through Learning}.
\newblock  \bibinfo{number}{arXiv:2410.04376} (\bibinfo{date}{Oct.} \bibinfo{year}{2024}).
\newblock
\urldef\tempurl%
\url{https://doi.org/10.48550/arXiv.2410.04376}
\showDOI{\tempurl}
\newblock
\shownote{arXiv:2410.04376 [cs].}


\bibitem[\protect\citeauthoryear{Johari, Li, Liskovich, and Weintraub}{Johari et~al\mbox{.}}{2022}]%
        {johari2022experimental}
\bibfield{author}{\bibinfo{person}{Ramesh Johari}, \bibinfo{person}{Hannah Li}, \bibinfo{person}{Inessa Liskovich}, {and} \bibinfo{person}{Gabriel~Y Weintraub}.} \bibinfo{year}{2022}\natexlab{}.
\newblock \showarticletitle{Experimental design in two-sided platforms: An analysis of bias}.
\newblock \bibinfo{journal}{\emph{Management Science}} \bibinfo{volume}{68}, \bibinfo{number}{10} (\bibinfo{year}{2022}).
\newblock


\bibitem[\protect\citeauthoryear{Kong, Yin, and Li}{Kong et~al\mbox{.}}{2022}]%
        {Kong_Yin_Li_2022}
\bibfield{author}{\bibinfo{person}{Fang Kong}, \bibinfo{person}{Junming Yin}, {and} \bibinfo{person}{Shuai Li}.} \bibinfo{year}{2022}\natexlab{}.
\newblock \showarticletitle{Thompson Sampling for Bandit Learning in Matching Markets}.
\newblock \bibinfo{journal}{\emph{ArXiv Preprint}} (\bibinfo{date}{May} \bibinfo{year}{2022}).
\newblock
\urldef\tempurl%
\url{http://arxiv.org/abs/2204.12048}
\showURL{%
\tempurl}
\newblock
\shownote{arXiv:2204.12048 [cs].}


\bibitem[\protect\citeauthoryear{Lee and Schwarz}{Lee and Schwarz}{2009}]%
        {Lee_Schwarz_2009}
\bibfield{author}{\bibinfo{person}{Robin Lee} {and} \bibinfo{person}{Michael Schwarz}.} \bibinfo{year}{2009}\natexlab{}.
\newblock \bibinfo{booktitle}{\emph{Interviewing in Two-Sided Matching Markets}}.
\newblock \bibinfo{address}{Cambridge, MA}. w14922 pages.
\newblock
\urldef\tempurl%
\url{https://doi.org/10.3386/w14922}
\showDOI{\tempurl}


\bibitem[\protect\citeauthoryear{Liu, Mania, and Jordan}{Liu et~al\mbox{.}}{2020}]%
        {centralized}
\bibfield{author}{\bibinfo{person}{Lydia~T. Liu}, \bibinfo{person}{Horia Mania}, {and} \bibinfo{person}{Michael~I. Jordan}.} \bibinfo{year}{2020}\natexlab{}.
\newblock \showarticletitle{Competing Bandits in Matching Markets}.
\newblock \bibinfo{journal}{\emph{arXiv:1906.05363 [cs, stat]}} (\bibinfo{date}{Jul} \bibinfo{year}{2020}).
\newblock
\urldef\tempurl%
\url{http://arxiv.org/abs/1906.05363}
\showURL{%
\tempurl}
\newblock
\shownote{arXiv: 1906.05363.}


\bibitem[\protect\citeauthoryear{Liu, Ruan, Mania, and Jordan}{Liu et~al\mbox{.}}{2021}]%
        {caucb}
\bibfield{author}{\bibinfo{person}{Lydia~T. Liu}, \bibinfo{person}{Feng Ruan}, \bibinfo{person}{Horia Mania}, {and} \bibinfo{person}{Michael~I. Jordan}.} \bibinfo{year}{2021}\natexlab{}.
\newblock \showarticletitle{Bandit Learning in Decentralized Matching Markets}.
\newblock \bibinfo{journal}{\emph{J. Mach. Learn. Res.}} \bibinfo{volume}{22}, \bibinfo{number}{1}, Article \bibinfo{articleno}{211} (\bibinfo{date}{jan} \bibinfo{year}{2021}), \bibinfo{numpages}{34}~pages.
\newblock
\showISSN{1532-4435}


\bibitem[\protect\citeauthoryear{Pagare and Ghosh}{Pagare and Ghosh}{2023}]%
        {pagare2023twosided}
\bibfield{author}{\bibinfo{person}{Tejas Pagare} {and} \bibinfo{person}{Avishek Ghosh}.} \bibinfo{year}{2023}\natexlab{}.
\newblock \showarticletitle{Two-Sided Bandit Learning in Fully-Decentralized Matching Markets}. In \bibinfo{booktitle}{\emph{ICML 2023 Workshop The Many Facets of Preference-Based Learning}}.
\newblock
\urldef\tempurl%
\url{https://openreview.net/forum?id=1oFkZbODgD}
\showURL{%
\tempurl}


\bibitem[\protect\citeauthoryear{Roth and Peranson}{Roth and Peranson}{1999}]%
        {roth1999redesign}
\bibfield{author}{\bibinfo{person}{Alvin~E Roth} {and} \bibinfo{person}{Elliott Peranson}.} \bibinfo{year}{1999}\natexlab{}.
\newblock \showarticletitle{The redesign of the matching market for {American} physicians: Some engineering aspects of economic design}.
\newblock \bibinfo{journal}{\emph{American economic review}} \bibinfo{volume}{89}, \bibinfo{number}{4} (\bibinfo{year}{1999}), \bibinfo{pages}{748--780}.
\newblock


\bibitem[\protect\citeauthoryear{Roth and Xing}{Roth and Xing}{1997}]%
        {Roth_Xing_1997}
\bibfield{author}{\bibinfo{person}{Alvin~E. Roth} {and} \bibinfo{person}{Xiaolin Xing}.} \bibinfo{year}{1997}\natexlab{}.
\newblock \showarticletitle{Turnaround Time and Bottlenecks in Market Clearing: Decentralized Matching in the Market for Clinical Psychologists}.
\newblock \bibinfo{journal}{\emph{Journal of Political Economy}} \bibinfo{volume}{105}, \bibinfo{number}{2} (\bibinfo{date}{Apr} \bibinfo{year}{1997}), \bibinfo{pages}{284–329}.
\newblock
\showISSN{0022-3808, 1537-534X}
\urldef\tempurl%
\url{https://doi.org/10.1086/262074}
\showDOI{\tempurl}


\bibitem[\protect\citeauthoryear{Sankararaman, Basu, and Abinav~Sankararaman}{Sankararaman et~al\mbox{.}}{2021}]%
        {sankararaman21}
\bibfield{author}{\bibinfo{person}{Abishek Sankararaman}, \bibinfo{person}{Soumya Basu}, {and} \bibinfo{person}{Karthik Abinav~Sankararaman}.} \bibinfo{year}{2021}\natexlab{}.
\newblock \showarticletitle{Dominate or Delete: Decentralized Competing Bandits in Serial Dictatorship}. In \bibinfo{booktitle}{\emph{Proceedings of The 24th International Conference on Artificial Intelligence and Statistics}} \emph{(\bibinfo{series}{Proceedings of Machine Learning Research}, Vol.~\bibinfo{volume}{130})}, \bibfield{editor}{\bibinfo{person}{Arindam Banerjee} {and} \bibinfo{person}{Kenji Fukumizu}} (Eds.). \bibinfo{publisher}{PMLR}, \bibinfo{pages}{1252--1260}.
\newblock
\urldef\tempurl%
\url{https://proceedings.mlr.press/v130/sankararaman21a.html}
\showURL{%
\tempurl}


\bibitem[\protect\citeauthoryear{Shah, Ferguson, and Marden}{Shah et~al\mbox{.}}{2024}]%
        {Shah_Ferguson_Marden_2024}
\bibfield{author}{\bibinfo{person}{Vade Shah}, \bibinfo{person}{Bryce~L. Ferguson}, {and} \bibinfo{person}{Jason~R. Marden}.} \bibinfo{year}{2024}\natexlab{}.
\newblock \showarticletitle{Learning Optimal Stable Matches in Decentralized Markets with Unknown Preferences}.
\newblock  \bibinfo{number}{arXiv:2409.04669} (\bibinfo{date}{Sept.} \bibinfo{year}{2024}).
\newblock
\urldef\tempurl%
\url{https://doi.org/10.48550/arXiv.2409.04669}
\showDOI{\tempurl}
\newblock
\shownote{arXiv:2409.04669 [cs, eess].}


\bibitem[\protect\citeauthoryear{Thompson}{Thompson}{1933}]%
        {original_thompson}
\bibfield{author}{\bibinfo{person}{William~R. Thompson}.} \bibinfo{year}{1933}\natexlab{}.
\newblock \showarticletitle{On the Likelihood that One Unknown Probability Exceeds Another in View of the Evidence of Two Samples}.
\newblock \bibinfo{journal}{\emph{Biometrika}} \bibinfo{volume}{25}, \bibinfo{number}{3/4} (\bibinfo{year}{1933}), \bibinfo{pages}{285--294}.
\newblock
\showISSN{00063444}
\urldef\tempurl%
\url{http://www.jstor.org/stable/2332286}
\showURL{%
\tempurl}


\bibitem[\protect\citeauthoryear{Zhang and Fang}{Zhang and Fang}{2024}]%
        {zhang2024decentralized}
\bibfield{author}{\bibinfo{person}{YiRui Zhang} {and} \bibinfo{person}{Zhixuan Fang}.} \bibinfo{year}{2024}\natexlab{}.
\newblock \showarticletitle{Decentralized Two-Sided Bandit Learning in Matching Market}. In \bibinfo{booktitle}{\emph{The 40th Conference on Uncertainty in Artificial Intelligence}}.
\newblock
\urldef\tempurl%
\url{https://openreview.net/forum?id=RUazoBVXIN}
\showURL{%
\tempurl}


\bibitem[\protect\citeauthoryear{Zhang, Wang, and Fang}{Zhang et~al\mbox{.}}{2022}]%
        {communicative}
\bibfield{author}{\bibinfo{person}{Yirui Zhang}, \bibinfo{person}{Siwei Wang}, {and} \bibinfo{person}{Zhixuan Fang}.} \bibinfo{year}{2022}\natexlab{}.
\newblock \showarticletitle{Matching in Multi-arm Bandit with Collision}. In \bibinfo{booktitle}{\emph{Advances in Neural Information Processing Systems}}, \bibfield{editor}{\bibinfo{person}{Alice~H. Oh}, \bibinfo{person}{Alekh Agarwal}, \bibinfo{person}{Danielle Belgrave}, {and} \bibinfo{person}{Kyunghyun Cho}} (Eds.).
\newblock
\urldef\tempurl%
\url{https://openreview.net/forum?id=Ixp6pznZgv7}
\showURL{%
\tempurl}


\end{thebibliography}

\newpage
\appendix
\pagenumbering{arabic}
\onecolumn
\section{Full Algorithm for OCA-UCB}

\begin{algorithm*}
    \caption{Optimistic Conflict Avoiding - UCB (OCA-UCB)}
    \label{alg:oca_ucb}
    \begin{tabular}{p{0.48\linewidth} p{0.48\linewidth}}
        \begin{algorithmic}[1]
        \State \textbf{Input:} $\lambda$, $H$
        \State \textbf{Output:} A stable match
        
        \For{$t \in \{1, \cdots , H\}$}
            \For{$i \in \{1, \cdots, N\}$}
                \If{$t == 1$}
                    \State $A_i(t) \gets a_k, k \sim \text{Uniform}(\left[ K\right])$
                \Else
                    \State $D_i(t) \sim Ber(\lambda)$ 
                    \If{$D_i(t) == 0$}
                        \State $A_i(t) \gets$ \Call{GetBestArm}{$p_i, t$}
                    \Else
                        \State $A_i(t) \gets A_i(t-1)$
                    \EndIf
                \EndIf
            \EndFor
        
            \For{$k \in \{1, \cdots, K\}$}
                \State $p_w \gets$ \Call{ResolveRequests}{$a_k, t$}
                \If{$p_w \neq \emptyset$}
                    \State $r_{p_w}, r_{a_k} \gets$ \smallcall{Sample-Rewards}{$p_w, a_k$}
                    \State \smallcall{Arm-Update-Belief}{$a_k, p_w, r_{a_k}$}
                    \State \smallcall{Player-Update-Belief}{$p_w, a_k, r_{p_w}$}
                    \State $L \gets \left\{p_j: A_j(t) = a_k \right\} \setminus \left\{p_w\right\}$
                    \For{$p_l \in L$}
                        \State \smallcall{UpdateConflictBelief}{$p_l, p_w, a_k$} 
                    \EndFor
                \EndIf
            \EndFor
        \EndFor
    \end{algorithmic}
        
        &
        
        \begin{algorithmic}[1]
            \resetlinenumber
            
            \Procedure{GetBestArm }{$p_i$, $t$}
            \State $S_i(t) \gets \emptyset$
            \For{$k \in \{1, \cdots, K\}$}
                \State $p_{\bar{p}} \gets$ $p_j \text{ s.t. } \bar{A}_j(t-1)=a_k$
                \If{$p_{\bar{p}} \in O_{(i)}^{k_l} \vee p_{\bar{p}}  = \emptyset$} 
                    \State $S_i(t) \gets S_i(t) \cup \{a_k\}$
                \EndIf
            \EndFor
            \State  $k \gets \underset{a_j \in S_i(t)}{\argmax} \left[ \UCB_i^j \right]$ 
            \State \Return $a_k$
            \EndProcedure
        
            \resetlinenumber

            \vspace{0.75cm}
           \Procedure{{\footnotesize UpdateConflictBelief}}{$p_i, p_j, a_k$}
                \State $O_{(i)}^{k_l} \gets O_{(i)}^{k_l} \setminus p_j$
                \State $O_{(i)}^{k_h} \gets O_{(i)}^{k_h} \cup \{p_j\}$
                \Statex \Comment{$p_i$ lost conflict with $p_j$ for $a_k$}
            \EndProcedure
            
        \end{algorithmic}
    \end{tabular}
\end{algorithm*}

\section{Full Proof of Convergence of OCA-UCB}
\label{appendix:oca_ucb_proof}

As Algorithm \ref{alg:oca_ucb} progresses, each of the players will have complete sets that represent the arms' true preferences with respect to the other players. In fact, we show theoretically that this approach to the solution does not affect the convergence guarantee of the CA-UCB algorithm in the APCK Model in Theorem \ref{thm:simple_theorem}.

\SimpleTheorem*

Recall some terminology. A triplet $Q_{ij}^k = (p_i, p_j,  a_k)$ is \emph{inconsistent} if $p_i$ believes $p_i \succ_{a_k} p_j$ but the opposite is true. Then, let us define a plausible set constructed by $p_i$ to be inconsistent with respect to $p_j$ and $a_k$ if $a_k \in S_i(t)$, denoted by $\hat{S}_{ij}^{k}(t)$. Finally, let us define an inconsistent plausible set as being \emph{inconsequential} at time $t$ if $\argmax_m \{\UCB_i^m : a_m \in S_i(t)\} \neq k$ i.e. $p_i$ does not attempt $a_k$ OR if $\argmax_m \{\UCB_j^m : a_m \in S_j(t)\} \neq k$ i.e. $p_j$ does not attempt $a_k$. We begin the proof of the convergence guarantee by first presenting two lemmas:  

\begin{lemma}
At each time-step, inconsistent sets $\hat{S}_{ij}^{k}(t)$ are either inconsequential or get resolved such that $S_i(t+1)$ is no longer inconsistent. 
\label{lemm:inconsistency_resolution}
\end{lemma}
\begin{proof}
Consider an inconsistent triplet $Q_{ij}^k = (p_i, p_j,  a_k)$ such that $\bar{A}_{j}(t-1) = a_k$. Then, at time $t$, $\hat{S}_{ij}^{k}(t)$- the plausible set constructed by $p_i$ will be inconsistent. One of three things must happen at time $t$ with regard to $\hat{S}_{ij}^{k}(t)$. (1) $p_j$ attempts $a_k$ either due to $\lambda$, or because $\argmax_m \{\UCB_j^m : a_m \in S_j(t)\} = k$. $p_i$ also attempts $a_k$ because $\argmax_m \{\UCB_i^m : a_m \in \hat{S}_{ij}^{k}(t) \} = k$. This leads to a  conflict, resulting in the inconsistency getting resolved via $p_i$ receiving feedback about matching information. (2) $p_j$ attempts $a_k$ but $p_i$ does not. This implies $\argmax_m \{\UCB_i^m : a_m \in \hat{S}_{ij}^{k}(t)\} \neq k$, which in turn implies $\hat{S}_{ij}^{k}(t)$ is inconsequential with respect to $p_j$ and $a_k$. (3) $p_j$ does not attempt $a_k$ but $p_i$ does. This implies $\hat{S}_{ij}^{k}(t)$ is inconsequential with respect to $p_j$ and $a_k$. However, two further distinct sub-cases follow from this. First, if $\bar{A}_i(t) = a_k$ then $p_i$ was the most preferred among $a_k$'s incoming pull requests. The plausible set constructed at $(t+1)$ is no longer inconsistent with respect to $p_j$ and $a_k$ because $a_k$ will be included in it by virtue of it being successfully pulled by $p_i$ at $t$, and not because of $p_i$'s belief $Q_{ij}^k$. And second, if $\bar{A}_i(t) \neq a_k$ then $a_k$ will not be included in plausible set constructed at $(t+1)$ as $p_i$ will lose a conflict with $M(a_k)$ and $p_i$'s belief will be updated with respect to $M(a_k)$ and $a_k$ if it has not already. 
\end{proof}

\begin{lemma}
Eventually all inconsistent plausible sets constructed are resolved or are inconsequential.
\label{lemm:inconsistency_resolution_all}
\end{lemma}
\begin{proof}
First, begin by noting that there are only finitely many inconsistent triplets in any given APKP model. Then, a simple recursive argument will suffice to show that each one gets resolved or becomes inconsequential. The first and second cases from Lemma \ref{lemm:inconsistency_resolution} are base cases. The first case corrects $p_i$ belief altogether, and in the second case the incorrect belief never gets used in any significant way. In the first sub-case of case three, if $a_k$ is in-fact $p_i$'s best possible achievable arm, then it will continue getting matched with $a_k$ regardless of it's wrong belief (inconsequential w.r.t $a_k$). Nonetheless, if $\exists p_j$ at $(t+1)$ such that triplet $Q_{ij}^k = (p_i, p_j, a_k)$ exists, then we can apply Lemma \ref{lemm:inconsistency_resolution} recursively. And finally, in the second sub-case of case three from Lemma \ref{lemm:inconsistency_resolution}, $p_i$ loses the conflict to some other player which implies that some belief gets updated resulting in the resolution of some triplet. However, if $\exists p_j$ at $(t+1)$ such that triplet $Q_{i,j}^k = (p_i, p_j, a_k)$ exists, then again we can apply Lemma \ref{lemm:inconsistency_resolution} recursively.
\end{proof}

The proof of Theorem \ref{thm:simple_theorem} is the immediate consequence of lemmas \ref{lemm:inconsistency_resolution}, and \ref{lemm:inconsistency_resolution_all}. Once all the inconsistent plausible sets are either resolved, or are inconsequential, the best arm picked from each player's plausible set will be the same as the one picked from CA-UCB's plausible set. This implies that this approach to updaing beliefs has no effect on CA-UCB algorithm's guarantee on convergence. \qed
\newpage
\section{Full Algorithm for PCA-SCA}
\label{appendix:pca_sca_full_aglorithm}

\begin{algorithm*}[h]
\caption{Probabilistic Conflict Avoiding - Simultaneous Choice Algorithm (PCA-SCA)}
\label{alg:pca_daa}
\begin{tabular}{p{0.48\linewidth} p{0.48\linewidth}}
    \begin{algorithmic}[1]
        \State \textbf{Input:} $\lambda$, $\kappa$, $H$
        \State \textbf{Output:} A stable match
        
        \For{$t \in \{1, \cdots , H\}$}
            \For{$i \in \{1, \cdots, N\}$}
                \If{$t == 1$}
                    \State $A_i(t) \gets a_k, k \sim \text{Uniform}(\left[ K\right])$
                \Else
                    \State $D_i(t) \sim Ber(\lambda)$ \label{line:bernoulli_sample}
                    \If{$D_i(t) == 0$}
                        \State $A_i(t) \gets$ \Call{GetBestArm}{$p_i, t$}
                    \Else
                        \State $A_i(t) \gets A_i(t-1)$
                    \EndIf
                \EndIf
            \EndFor
        
            \For{$k \in \{1, \cdots, K\}$}
                \State $p_w \gets$ \Call{ResolveRequests}{$a_k, t$}
                \If{$p_w \neq \emptyset$}
                    \State $r_{p_w}, r_{a_k} \gets$ \smallcall{Sample-Rewards}{$p_w, a_k$}
                    \State \smallcall{Arm-Update-Belief}{$a_k, p_w, r_{a_k}$}
                    \State \smallcall{Player-Update-Belief}{$p_w, a_k, r_{p_w}$}
                    \State $L \gets \left\{p_j: A_j(t) = a_k \right\} \setminus \left\{p_w\right\}$
                    \For{$p_l \in L$}
                        \State \smallcall{Update-Win-Probability}{$p_l, p_w, a_k$} \label{line:update_probability}
                    \EndFor
                \EndIf
            \EndFor
        \EndFor
    \end{algorithmic}
&
    \begin{algorithmic}[1]
    \resetlinenumber
        {\footnotesize
        \Procedure{GetBestArm }{$p_i$, $t$}
            \For{$k \in \{1, \cdots, K\}$}
                \State $p_{\bar{p}} \gets$ $p_j \text{ s.t. } \bar{A}_j(t-1)=a_k$
                \State {\tiny REWARDS$[k] \gets \mathcal{R}_i^k$} \Comment{{\tiny Reward estimate}} \label{line:optimism}
                \State {$P[k] \gets f\left( \hat{P}_{i \bar{p}}^k\right)$} \Comment{{\tiny Optimism Function}} \label{line:optimism}
            \EndFor
            \State  $j \gets \arg \max \left( \text{REWARDS} \circ \text{P} \right)$ \label{line:circ}
            \State \Return $a_j$
        \EndProcedure
        }

    \resetlinenumber
        {\footnotesize
        \Procedure{ResolvePullRequests}{$a_k, t$} \label{proc:resolve_pull_requests}
            \State $A_{rq} \gets \left\{p_i : A_i(t)  = a_k \right\}$\Comment{{\tiny pull requests}}
            \State $O_{inf} \gets \emptyset$ \Comment{{\tiny players with $\infty$ reward estimate}}
            \State $O_{bst} \gets \emptyset$ \Comment{{\tiny players with best overlapping reward estimate}}
            \State $I_L, I_U \gets \infty, - \infty$  \Comment{{\tiny Keep track of best reward interval}}
            \For{$p_i \in A_{rq}$}
                \State $\mathcal{R}_i^U \gets {\tiny \text{Upper bound on reward estimate for } p_i}$ \label{line:upper}
                \State $\mathcal{R}_i^L \gets {\tiny \text{Lower bound on reward estimate for } p_i}$ \label{line:lower}
                \If{$\mathcal{R}_i^U = \infty$}
                    \State $O_{inf} = O_{inf} \cup \{p_i\}$
                    \State {\footnotesize \textbf{Continue} }
                \EndIf
                \If {$\mathcal{R}_i^L > I_U$} \Comment{{\tiny Better interval}}
                    \State $I_H \gets \mathcal{R}_i^U $, $I_L \gets \mathcal{R}_i^L $
                    \State $O_{bst} = \{p_i\}$
                \Else{$(I_U \leq \mathcal{R}_i^L) \wedge (\mathcal{R}_i^U  \leq I_L)$ } \Comment{{\tiny Overlap}} \label{line:check_overlap}
                    \State $O_{bst} = O_{bst} \cup \{p_i\}$
                \EndIf
            \EndFor
            \If {$O_{inf} \neq \emptyset$}
                \State \Return $\sim \text{Uniform}(O_{inf})$ \label{line:infinite_players}
            \Else 
                \State \Return $\sim \text{Uniform}(O_{bst})$ \label{line:best_players}
            \EndIf
        \EndProcedure}
    \end{algorithmic}
\end{tabular}
\end{algorithm*}

\section{Full Proof of Convergence of PCA-UCB}
\label{proof:full}

In this section, we re-establish all the necessary information for the proof's context. We consider the APU model with $K$ arms and $N$ players.

\subsection{Preliminaries for the Arms}
Each arm $k$ has a true mean reward associated with player $i$ denoted by $\mu_i^k$, and let $\hat{\mu}_i^k(t)$ denote the empirical mean computed from $n_i^k(t)$ samples by time $t$. Then, the corresponding confidence intervals that arm $k$ keeps track of are:
\begin{equation}\
    \begin{array}{rcl}
        \LCB_i^k(t) &=& \hat{\mu}_i^k(t) - c_i^k(t), \\
        \UCB_i^k(t) &=& \hat{\mu}_i^k(t) + c_i^k(t)
    \end{array}
    \label{eqn:cnfdnce}
\end{equation}
\noindent where 
\[
c_i^k(t) = \sqrt{\frac{3 \ln(t)}{2n_i^k(t)}}.
\]
Also, for any pair of players $p_i$ and $p_j$, define:
\begin{equation*}
    \begin{array}{rcl}
        \Delta_{ij}^k & = & \mu_i^k - \mu_j^k \quad \text{for } \mu_i^k > \mu_j^k, \\
        \DeltaMin & =& \underset{i \neq j, \forall k}{\min} \lvert\mu_i^k - \mu_j^k \rvert, \\
        n_{ij}^k(t) &=& \min\{n_i^k(t), n_j^k(t)\}.
    \end{array}
\end{equation*}
Note that, based on our construction, $\DeltaMin = 1$.

\subsection{Preliminaries for the Players}
Each player \(p_i\) maintains an empirical win probability \(\hat{P}_{ij}^k(t)\) against player \(p_j\) for arm \(a_k\), where the true probability \(P_{ij}^k\in\{0,1\}\) is unknown. Let \(N_{c_{ij}}^k(T)\) denote the number of conflicts between \(p_i\) and \(p_j\) for arm \(a_k\) up to time \(T\). We use the capital \(N_c\) since our proof considers the period before the arms’ confidence intervals become disjoint (at time \(T\)), at which point the number of collisions is fixed.

\subsubsection{Conflict Feedback to the Players}
In our analysis, we assume without loss of generality that players are ordered so that $\mu_i^k > \mu_j^k$ when $p_i \succ_{a_k} p_j$. We say that the confidence intervals for arm $a_k$ \textbf{are \emph{disjoint}} if, for every pair $\{p_i, p_j\}$ with $\mu_i^k > \mu_j^k$,
\begin{equation*}
\LCB_i^k(t) > \UCB_j^k(t).
\end{equation*}
Conversely, if for some $\{p_i, p_j\}$ with $\mu_i^k > \mu_j^k$ we have
\begin{equation*}
\LCB_i^k(t) \leq \UCB_j^k(t),
\end{equation*}
then the confidence intervals for those two players are said to be \textbf{overlapping}. When an arm's confidence intervals overlap for a set of players, the conflict feedback is provided uniformly at random. Specifically, the arm selects the proposal with the highest $\UCB$ estimate; if there is any overlap with another player’s interval, it chooses uniformly at random from the contenders. As a result, the win probability estimates for losing players concentrate around $0.5$. Once the intervals become disjoint, however, the feedback becomes deterministic.

\begin{definition}[Deterministic Feedback Event $\alpha(t)$]
\label{def:alpha_t}
For any time $t$, define
\begin{equation*}
    \alpha(t) = \Bigl\{ \forall\, k\in[K],\, \forall\, i,j \text{ with } \mu_i^k > \mu_j^k:\; \LCB_i^k(t) > \UCB_j^k(t) \Bigr\}.
\end{equation*}
Thus, $\alpha(t)$ is the event that, for every arm, the players’ confidence intervals are disjoint. When $\alpha(t)$ holds, the arm’s feedback is deterministic.
\end{definition}

\subsubsection{Arm Choice Mechanism of Players}
Players use an \textbf{\textit{optimism function}} to choose an arm at each time step. The function is defined as:

\begin{equation}
  \tag{\ref{eqn:optimism_fxn}}
  \optimismfxn
\end{equation}

\noindent where $\kappa > 0$ is the \textbf{\textit{optimism parameter}}. By choosing $\kappa$ sufficiently large, even if a player’s raw empirical win estimate is as low as $0.5-\epsilon$ (for some $\epsilon>0$), the adjusted value $f(x)$ is boosted above a desired threshold, ensuring the arm continues to be selected. (This is formalized later in one of our lemmas.)

\subsubsection{Plausible Sets and Inconsistent Triplets}
In CA-UCB and OCA-UCB, for a player $p_i$, the \emph{plausible set} at time-step $t$ is defined as
\begin{equation*}
S^{(i)}(t) := \{a_j : p_i \succ_{a_j} p_k, \text{ where } \bar{A}^k(t-1) = a_j\}.
\end{equation*}
That is, $S^{(i)}(t)$ is the set of all arms $a_j$ for which, at time-step $t-1$, some other player $p_k \neq p_i$ successfully pulled the arm, yet $p_i$ \textbf{knows} that the arm’s inherent preference favors $p_i$ over $p_k$. In CA-UCB, players had direct access to arms’ preferences, whereas in OCA-UCB, the arms’ preferences were learned deterministically. In the context of the APKP model, one may view an arm $a_k \in S^{(i)}(t)$ as having a win probability of 1 (with all other arms having a win probability of 0), thereby considering arms with a win probability of 1 as “plausible”. In the APU model, however, plausibility must be interpreted more generally as any arm with a non-zero probability of being won.

\begin{definition}[Inconsistent Triplet]
\label{dfn:inconsistent_triplet}
    We now define an \emph{inconsistent triplet} in the APU model. Let $P_{ij}^k$ denote the true probability of player $p_i$ winning against player $p_j$ in a conflict for arm $a_k$, and let $\hat{P}_{ij}^k$ be the corresponding empirical probability. The triplet $(p_i, p_j, a_k)$ is said to be \emph{inconsistent} if
        \begin{equation*}
            \Bigl[ \Bigl\{ \Bigl(P_{ij}^k =1 \Bigr) \wedge \Bigl( \hat{P}_{ij}^k \leq 1 - \epsilon \Bigr) \Bigr\} \vee  \Bigl\{ \Bigl(P_{ij}^k = 0 \Bigr) \wedge \Bigl( \hat{P}_{ij}^k \geq \epsilon \Bigr) \Bigr\} \Bigr],
        \end{equation*}
\end{definition}

\noindent i.e., the player’s empirical belief about its win probability in the conflict deviates significantly (by at least $\epsilon$) from the true outcome—either overestimating a win or underestimating a loss. Next, we define a \emph{inconsequential triplet}

\begin{definition}[Inconsequential Triplet]
    \label{dfn:inconsequantial_triplet}
    An inconsistent triplet is considered \emph{inconsequential} if there exists an alternative arm that offers a better reward prospect. Formally, consider two triplets $(p_i, p_j, a_k)$ and $(p_i, p_j', a_k')$. If, at time $t$, the triplet $(p_i, p_j, a_k)$ is inconsistent but $\UCB_i^{k'} > \UCB_i^{k}$ and the triplet $(p_i, p_j', a_k')$ is consistent (i.e., arm $a_{k'}$ is a better prospect for $p_i$), then player $p_i$ will select $a_{k'}$. In this case, the inconsistency in $(p_i, p_j, a_k)$ is rendered inconsequential.
\end{definition}
\noindent And finally a triplet is considered \emph{consequential} if it is not inconsequential.

\subsection{Main Results}

The main proof can be summarized in the following main theorems: 

\TheoremOne*

\bigskip

\TheoremTwo*

\bigskip

\TheoremThree*

\subsection{Proof of Theorem \ref{thm:1}}

We will begin the proof of Theorem \ref{thm:1} by first establishing some important lemmas that relate to bounding the time that is taken for the arms to learn their preferences (i.e. the event $\alpha(t)$ happening). We begin by looking at the tracked reward estimates by an arm $a_k$ for a single pair of players $p_i$ and $p_j$.

\begin{lemma}[Single Pair of Players Overlap] Let $p_i$ and $p_j$ be two players such that $\mu_i^k > \mu_j^k$ and $\hat{\mu}_i^k$ and $\hat{\mu}_j^k$ be the empirical reward estimate by arm $a_k$ for the players respectively at time $t$. With the confidence intervals are defined as in Equation \ref{eqn:cnfdnce}, the probability that the confidence intervals for $p_i$ and $p_i$ overlap is bounded by: 
    \begin{equation*}
        P\left( \LCB_i^k(t) \leq \UCB_j^k(t) \right) \leq \exp\left( -\frac{ \left( \Delta_{ij}^k - (c_i^k(t) + c_j^k(t)) \right)^2 }{ 2 \left( \frac{1}{n_i^k(t)} + \frac{1}{n_j^k(t)} \right) } \right)
    \end{equation*}
where $ \Delta_{ij}^k = \mu_i^k - \mu_j^k $ is the true mean difference between players $ p_i $ and $ p_j $, and $ n_i^k(t) $ and $ n_j^k(t) $ are the numbers of samples for each player at time $ t $.
\label{lemma:single_pair_arm}
\end{lemma}

\begin{proof} In the context of this proof, it is clear that we are talking about arm $a_k$, so we will not use the superscript of $k$ for clarity. The event $ \LCB_i(t) \leq \UCB_j(t) $ can be rewritten as:
        $$\hat{\mu}_i(t) - \hat{\mu}_j(t) \leq c_i(t) + c_j(t)$$
    We have that $ \Delta_{ij} = \mu_i - \mu_j $, which is positive since $ \mu_i > \mu_j $. Then we can express the difference in sample means as:
        $$\hat{\mu}_i(t) - \hat{\mu}_j(t) = \Delta_{ij} + \left( \hat{\mu}_i(t) - \mu_i \right) - \left( \hat{\mu}_j(t) - \mu_j \right)$$
    Thus, the event $ \LCB_i(t) \leq \UCB_j(t) $ becomes:
        $$\left( \hat{\mu}_i(t) - \mu_i \right) - \left( \hat{\mu}_j(t) - \mu_j \right) \leq c_i(t) + c_j(t) - \Delta_{ij}$$
    Define $ \epsilon_{ij} = \Delta_{ij} - (c_i(t) + c_j(t)) $. Then we need to bind the probability of the event
        $$Z = \left( \hat{\mu}_i(t) - \mu_i \right) - \left( \hat{\mu}_j(t) - \mu_j \right) \leq -\epsilon_{ij}$$
        
    \noindent We will use the Hoeffding Inequality for Sub-Gaussian random variables to do this. But first, we need to show that $Z$ is \textbf{Sub-Gaussian}. We do this in the following lemma. 
    
    \begin{sublemma}
    Let $\{X_{i,k}\}_{k=1}^{n_i(t)}$ be independent samples drawn from $\mathcal{N}(\mu_i,1)$ and $\{X_{j,k}\}_{k=1}^{n_j(t)}$ be independent samples drawn from $\mathcal{N}(\mu_j,1)$. Define the sample means
    $$ \hat{\mu}_i(t) = \frac{1}{n_i(t)} \sum_{k=1}^{n_i(t)} X_{i,k} \quad \text{and} \quad \hat{\mu}_j(t) = \frac{1}{n_j(t)} \sum_{k=1}^{n_j(t)} X_{j,k},$$
    and consider the random variable
    $$Z = \Bigl( \hat{\mu}_i(t) - \mu_i \Bigr) - \Bigl( \hat{\mu}_j(t) - \mu_j \Bigr).$$
    Then, $Z$ is a sub-Gaussian random variable with a sub-Gaussian parameter
    $$\sigma_Z = \sqrt{\frac{1}{n_i(t)} + \frac{1}{n_j(t)}},$$
    \label{sublemma:sub_gaussian}
\end{sublemma}
    
\begin{proof}
By properties of the normal distribution, we have:
    $$ \hat{\mu}_i(t) \sim \mathcal{N}\!\left(\mu_i, \frac{1}{n_i(t)}\right) \quad \text{and} \quad \hat{\mu}_j(t) \sim \mathcal{N}\!\left(\mu_j, \frac{1}{n_j(t)}\right).$$
And, we have that $Z$ is a random variable with: 
    $$ Z = \Bigl( \hat{\mu}_i(t) - \mu_i \Bigr) - \Bigl( \hat{\mu}_j(t) - \mu_j \Bigr).$$
Since,   
    $$ \hat{\mu}_i(t)-\mu_i \sim \mathcal{N}\!\left(0, \frac{1}{n_i(t)}\right) \quad \text{and} \quad \hat{\mu}_j(t)-\mu_j \sim \mathcal{N}\!\left(0, \frac{1}{n_j(t)}\right)$$
are independent, it follows that: 
    $$ Z \sim \mathcal{N}\!\left(0, \frac{1}{n_i(t)} + \frac{1}{n_j(t)}\right).$$
Let, 
    $$ \sigma_Z^2 = \frac{1}{n_i(t)} + \frac{1}{n_j(t)},$$
so that 
    $$ \sigma_Z = \sqrt{\frac{1}{n_i(t)} + \frac{1}{n_j(t)}}.$$
Recall that a random variable $X$ is said to be \emph{sub-Gaussian} with parameter $\sigma$ if
    $$\mathbb{E}\!\left[e^{t(X - \mathbb{E}[X])}\right] \le e^{\frac{\sigma^2 t^2}{2}} \quad \text{for all } t \in \mathbb{R}.$$
A well-known fact is that if $X \sim \mathcal{N}(0,\sigma^2)$, then
    $$ \mathbb{E}\!\left[e^{tX}\right] = e^{\frac{\sigma^2 t^2}{2}} \quad \text{for all } t \in \mathbb{R},$$
which shows that $X$ is sub-Gaussian with parameter $\sigma$. Since $Z \sim \mathcal{N}(0,\sigma_Z^2)$, we conclude that $Z$ is sub-Gaussian with parameter
    $$\sigma_Z = \sqrt{\frac{1}{n_i(t)} + \frac{1}{n_j(t)}},$$
\end{proof}

    \noindent Lemma \ref{sublemma:sub_gaussian} establishes $Z$ as a sub-Gaussian random variable. Then, using \textbf{the Hoeffding Inequality for Sub-Gaussian Random Variables} yields: 
        $$P(Z \leq -\epsilon) \leq \exp\left( -\frac{\epsilon^2}{2 \sigma^2} \right)$$
    Applying this to our event, we have
        $$P\left( \LCB_i(t) \leq \UCB_j(t) \right) \leq \exp\left( -\frac{ \epsilon_{ij}^2 }{ 2 \sigma_{ij}^2 } \right)$$
    Since $ \epsilon_{ij} = \Delta_{ij} - (c_i(t) + c_j(t)) $ and $ \sigma_{ij}^2 = \frac{1}{n_i(t)} + \frac{1}{n_j(t)} $, we obtain
        $$P\left( \LCB_i(t) \leq \UCB_j(t) \right) \leq \exp\left( -\frac{ \left( \Delta_{ij} - (c_i(t) + c_j(t)) \right)^2 }{ 2 \left( \frac{1}{n_i(t)} + \frac{1}{n_j(t)} \right) } \right)$$
\end{proof}

Now that we have established the probability bound for any given pair of players for an arm, we will extend this argument to all arms keeping track of the confidence intervals of all players. 

\begin{lemma}[All Pair of Players Overlap] 

    Consider $a_k$ be a fixed arm and $\{p_i\}_{i=1}^N$ be the set of players each with true means $\{\mu_i^k\}_{i=1}^N$. Let $\DeltaMin^k = \min \{\mu_i^k - \mu_j^k : \mu_i^k > \mu_j^k\}$, then using $n^k (t)= \underset{i}{\min}\left[n_i^k(t)\right]$ and the confidence interval $c^k(t) = \sqrt{\frac{3 \ln(t)}{2n^k(t)}}$, the probability that any pair of confidence intervals overlaps for arm $a_k$ satisfies:

        \begin{equation*}
            P\left( \bigcup_{i \neq j, \mu_i^k > \mu_j^k} \left\{ \LCB_i^k(t) \leq \UCB_j^k(t) \right\} \right) \leq \frac{N^2}{2} \cdot \exp\left( -\frac{ n^k \left( \Delta_{\text{min}}^k - 2 c^k(t) \right)^2 }{ 4 } \right)
        \end{equation*}
    
    \label{lemma:all_pairs_arm}
\end{lemma}

\begin{proof}
    In Lemma \ref{lemma:single_pair_arm}, we have established that for an arm $a_k$ and any two players $p_i$ and $p_j$ s.t. $\mu_i^k > \mu_j^k$: 

    $$P\left( \LCB_i^k(t) \leq \UCB_j^k(t) \right) \leq \exp\left( -\frac{ \left( \Delta_{ij}^k - (c_i^k(t) + c_j^k(t)) \right)^2 }{ 2 \left( \frac{1}{n_i^k(t)} + \frac{1}{n_j^k(t)} \right) } \right)$$

    \noindent Notice that if we let $\DeltaMin^k = \min \{\mu_i^k - \mu_j^k : \mu_i^k > \mu_j^k\}$ and set $n_k (t)= \underset{i}{\min}\left[n_i^k(t)\right]$, we have that: 

    \begin{itemize}
        \item $\Delta_{ij}^k \geq \DeltaMin^k$,
        \item $(c_i^k(t) + c_j^k(t)) \leq 2 c^k(t)$ where $c^k(t) = \sqrt{\frac{3 \ln(t)}{2n^k(t)}}$, and 
        \item $\left( \frac{1}{n_i^k(t)} + \frac{1}{n_j^k(t)} \right) \leq \frac{2}{n^k(t)}$
    \end{itemize}

    \noindent The following inequality still holds: 
    $$P\left( \LCB_i^k(t) \leq \UCB_j^k(t) \right) \leq \exp\left( - \frac{n^k(t) \left( \DeltaMin^k - 2 c^k(t) \right)^2}{4}\right)$$

     \noindent Next, the total number of players where $ \mu_i > \mu_j $ is at most $ \frac{N(N - 1)}{2} $. For each arm, we have $ N $ choices for player $ i $ and $ N $ choices for player $ j $, giving a total of $ N \times N = N^2 $ ordered pairs $ (i, j) $ where $ i $ and $ j $ are any two players (including the cases where $ i = j $). Since we are interested only in pairs $ (i, j) $ where $ i \neq j $, we exclude the $ N $ pairs where $ i = j $ (e.g., $ (1, 1), (2, 2), \dots, (N, N) $). Thus, there are $ N^2 - N = N(N - 1) $ pairs with $ i \neq j $. For each distinct pair $ (i, j) $ where $ i \neq j $, only one of the following statements can be true: either $ \mu_i > \mu_j $ or $ \mu_j > \mu_i $. Therefore, we only consider one direction for each pair (e.g., $ (i, j) $ and not $ (j, i) $), effectively halving the number of pairs. Since we only count each distinct pair $ (i, j) $ once, the number of pairs with $ i \neq j $ is halved. This gives us $ \frac{N(N - 1)}{2} $ pairs where we consider $ \mu_i > \mu_j $ for each distinct player pair $ (i, j) $. \\

     \noindent Finally, with a straightforward application of the union bound over all possible pairs, we get: 
     \begin{align*}
         P\left( \bigcup_{i \neq j, \mu_i^k > \mu_j^k} \left\{ \LCB_i^k(t) \leq \UCB_j^k(t) \right\} \right) & \leq \frac{N(N-1)}{2} \cdot \exp\left( -\frac{ n^k \left( \Delta_{\text{min}}^k - 2 c^k(t) \right)^2 }{ 4 } \right) \\
         & \leq \frac{N^2}{2} \cdot \exp\left( -\frac{ n^k \left( \Delta_{\text{min}}^k - 2 c^k(t) \right)^2 }{ 4 } \right)
     \end{align*}
\end{proof}

\begin{lemma}[All Arms for All Pair of Players Overlap]

    For $K$ arms with independent confidence interval calculations, the probability of any overlap on any arm satisfies: 

    \begin{equation*}
        P\left(\exists a_k \text{ with } \mu_i^k > \mu_j^k \text{ and } \LCB_i^k(t) \leq UCB_j^k(t)\right) \leq K \cdot \frac{N^2}{2} \cdot \exp\left( -\frac{ n(t) \left( \Delta_{\text{min}} - 2 c(t) \right)^2 }{ 4 } \right)
    \end{equation*}
    \noindent where $n(t) = \underset{k \in K}{\min}\left\{n^k(t)\right\}$, $\DeltaMin = \underset{k \in K}{\min}\left\{\DeltaMin^k\right\}$ and $c(t) = \sqrt{\frac{3 \ln(t)}{2n(t)}}$
    \label{lemma:all_arms_pairs_arm}
\end{lemma}

\begin{proof}
    The proof of this lemma is a straightforward application of Lemma \ref{lemma:all_pairs_arm} using the union bound over all $K$ arms and recognizing that replacing $n(t), \DeltaMin, \text{ and } c(t)$ in the expression still maintains the upper-bound on the probability. 
\end{proof}

\subsubsection{Putting it together}
Now, we will use Lemma \ref{lemma:all_arms_pairs_arm} to prove Theorem \ref{thm:1}. Notice that the probability that we have upper-bounded in Lemma \ref{lemma:all_arms_pairs_arm} is the complement of the event $\alpha(t)$ (i.e. the confidence intervals for all the players tracked by all the arms are disjoint). Hence: 
\begin{align*}
    P\left( \alpha(t) \text{ occurs at time } t \right) &= 1 -  P\left(\exists a_k \text{ with } \mu_i^k > \mu_j^k \text{ and } \LCB_i^k(t) \leq UCB_j^k(t)\right)\\
    & \leq  1 - \underbrace{K \cdot \frac{N^2}{2} \cdot \exp\left( -\frac{ n(t) \left( \Delta_{\text{min}} - 2 c(t) \right)^2 }{ 4 } \right)}_\delta
\end{align*}
\qed

\noindent \textbf{Important: }Notice that as $t \to \infty$, $n(t) \to \infty$. Since the contents inside the brackets are a positive number by being squared, the overall exponent $\to - \infty \implies \delta \to 0$. This will be important later in establishing that PCA-SCA's execution is identical to CA-UCB as $t \to \infty$.

\subsection{Proof of Theorem \ref{thm:2}}

Assume that the event $\alpha$ occurs at time $T$, so that the conditions of Theorem \ref{thm:1} are satisfied. Our goal is to analyze the evolution of the players’ empirical winning probabilities, $\hat{P}_{ij}^k$, both before and after $\alpha(T)$.

We begin by considering $\hat{P}_{ij}^k$ prior to $\alpha(T)$. This is formalized in the following lemma:

\begin{lemma}
    \label{lemma:probability_concentration_0.5}
    Assume event $\alpha$ has happened by time $T$. The empirical win estimate tracked by a player $p_i$ against player $p_j$ for an arm $a_k$ at time $T$ satisfies: 
    \begin{equation*}
        P(\hat{P}_{ij}^k \leq 0.5 - \epsilon) \leq exp \left( -2 N_{c_{ij}}^k(T) \epsilon^2 \right)
    \end{equation*}
    \noindent for some $\epsilon \geq 0$ and where $N_{c_{ij}}^k(T)$ is the the number of times players $p_i$ and $p_j$ have conflicted for arm $a_k$ up to time $T$.  
\end{lemma}

\begin{proof}
    The proof is a standard application of concentration of sum of i.i.d Bernoulli random samples. We know by the construction of the algorithm, while there is still overlap between the confidence interval of the arm-tracked reward for the players (i.e. until $\alpha$ has happened), the arm picks the players uniformly at random from the overlapping players. Then: 
    \begin{equation*}
        \hat{P}_{ij}^k = \frac{1}{N_{c_{ij}}^k(T)}\sum_{i=1}^{N_{c_{ij}}^k(T)}x_i \left(\text{ where } x_i \sim \text{Bernoulli}(0.5)\right)
    \end{equation*}
    By standard concentration, we have that for any $\epsilon > 0$: 
    \begin{equation*}
        P(\hat{P}_{ij}^k \leq 0.5 - \epsilon) \leq exp \left( -2 N_{c_{ij}}^k(T) \epsilon^2 \right)
    \end{equation*}
\end{proof}

Lemma \ref{lemma:probability_concentration_0.5} quantifies how unlikely it is for a player to have an empirical win estimate significantly below 0.5, which is crucial to prevent a player from discarding an arm due to an erroneously low conflict win estimate.

To enforce this safeguard, we employ the \emph{optimism function} defined in Equation \ref{eqn:optimism_fxn}, yielding the next lemma:

\begin{lemma}[Optimistic Picking of Arm by Player]
    \label{lemma:optimism}
    Let event $\alpha$ occur at time $T$. Suppose player $p_i$ has an empirical win estimate $\hat{P}_{ij}^k$ for winning a conflict against player $p_j$ on arm $a_k$. Define the \textbf{\textit{optimism function}} $f\colon [0,1] \to [0,1]$ by
    \begin{equation*}
        f(x)=
        \begin{cases}
        \displaystyle \frac{1-e^{-\kappa x}}{1-e^{-\kappa/2}}, & \text{if } 0\le x\le 0.5,\\[1ex]
        1, & \text{if } x>0.5.
        \end{cases}
    \end{equation*}
    \noindent Then for any $\epsilon>0$, there exists a sufficiently large constant $\kappa$ such that when $p_i$ uses $f\bigl(\hat{P}_{ij}^k\bigr)$ to select an arm, it will choose arm $a_k$ with probability at least
    \begin{equation*}
        1 - \exp\Bigl(-2\, N_{c_{ij}}^k(T)\, \epsilon^2\Bigr),
    \end{equation*}
    \noindent where $N_{c_{ij}}^k(T)$ denotes the number of conflicts between $p_i$ and $p_j$ up to time $T$.
\end{lemma}

\begin{proof}
    Let us pick a threshold $\epsilon > 0$ such that even if $\hat{P}_{ij}^k = 0.5 - \epsilon$, we have that $f(0.5 - \epsilon) \geq q > 0$, ensuring that the players' win heuristic remains sufficiently large to keep attempting the arm: 
    \begin{equation*}
        f(0.5-\epsilon)=\frac{1-e^{-\kappa (0.5-\epsilon)}}{1-e^{-\kappa/2}} \geq q
    \end{equation*}
    \noindent This inequality can be solved for $\kappa$ and any desired value of $\epsilon$ and $q$. Suppose then we pick a suitable $\kappa$, the player only fails to try the arm if $\hat{P}_{ij}^k < 0.5 - \epsilon$. By Lemma \ref{lemma:probability_concentration_0.5} we know that $P(\hat{P}_{ij}^k < 0.5 - \epsilon) \leq \exp\Bigl(-2\, N_{c_{ij}}^k(T)\, \epsilon^2\Bigr)$. Therefore, with probability at least $1 - \exp\Bigl(-2\, N_{c_{ij}}^k(T)\, \epsilon^2\Bigr)$, the player's win estimate will be at least $0.5 - \epsilon$, ensuring that the player continues to attempt the arm. 
\end{proof}

\noindent \textbf{Important: } Notice that, regardless of what value of $\epsilon$ we pick, as $\kappa \to \infty, f (0.5 - \epsilon) \to 1$ i.e. regardless of where we pick the threshold or how far the empirical estimate is from 0.5, we can always pick $\kappa$ to be arbitrarily large to reach a level to optimism required that the players attempt arms. 

We then extend this result to all possible triplets to ensure that no player stops prematurely due to adverse fluctuations:

\begin{lemma}
\label{lemma:optimism_all}
    Assume that event $\alpha$ occurs at time $T$ and that all players employ the \emph{optimism function} when selecting arms. Define 
        \begin{equation*}
        N_c(T) = \min_{\substack{i,j\in N\\k\in K}} \{N_{c_{ij}}^k(T)\},
        \end{equation*}
    the minimum number of conflicts observed among any pair of players on any arm by time $T$. Then, for any $\epsilon > 0$, the probability that all players continue attempting arms is at least
        \begin{equation*}
        1 - \frac{KN^2}{2} \exp\Bigl(-2\, N_c(T)\, \epsilon^2\Bigr).
        \end{equation*}
\end{lemma}

\begin{proof}
    By Lemma~\ref{lemma:probability_concentration_0.5}, for any players $p_i$ and $p_j$ and any arm $a_k$ (with $p_i \succ_{a_k} p_j$), we have
        \begin{equation*}
        P\Bigl(\hat{P}_{ij}^k \le 0.5 - \epsilon\Bigr) \le \exp\Bigl(-2\, N_{c_{ij}}^k(T)\, \epsilon^2\Bigr).
        \end{equation*}
    Since $N_{c_{ij}}^k(T) \ge N_c(T)$ for all such triplets, it follows that
        \begin{equation*}
        P\Bigl(\hat{P}_{ij}^k \le 0.5 - \epsilon\Bigr) \le \exp\Bigl(-2\, N_c(T)\, \epsilon^2\Bigr).
        \end{equation*}
    There are at most $K \cdot \frac{N(N-1)}{2}$ triplets in consideration in the system. Applying the union bound, the probability that \emph{any} triplet has an empirical win estimate below $0.5-\epsilon$ is at most
        \begin{equation*}
        \frac{KN^2}{2} \exp\Bigl(-2\, N_c(T)\, \epsilon^2\Bigr).
        \end{equation*}
    Finally, by Lemma~\ref{lemma:optimism}, as long as every empirical win estimate exceeds $0.5-\epsilon$, the optimism function guarantees that players will continue selecting arms. Thus, the probability that all players continue attempting arms is at least
        \begin{equation*}
        1 - \frac{KN^2}{2} \exp\Bigl(-2\, N_c(T)\, \epsilon^2\Bigr).
        \end{equation*}
\end{proof}

These lemmas establish a lower bound on the probability that players continue selecting the arms after $\alpha(T)$. Next, we set a suitable $\epsilon$ so that the tail probabilities align. Let $B$ denote the event that no triplet has an empirical win estimate $\hat{P}$ falling below $0.5-\epsilon$ after $\alpha(T)$. We require that
    \begin{equation*}
        P(B) \geq P(\alpha),
    \end{equation*}
    
    \begin{equation*}
        P(\alpha) \geq 1 - K \cdot \frac{N^2}{2} \exp\!\left( -\frac{ n(T) \left( \Delta_{\text{min}} - 2 c(T) \right)^2 }{ 4 } \right)
    \end{equation*}

    \begin{equation*}
        P(B) \geq 1 - \frac{KN^2}{2} \exp\!\Bigl(-2\, N_c(T)\, \epsilon^2\Bigr).
    \end{equation*}
Matching the tail probabilities yields
    \begin{equation*}
        \exp\!\left( -\frac{ n(T) \left( \Delta_{\text{min}} - 2 c(T) \right)^2 }{ 4 } \right) \approx \exp\!\Bigl(-2\, N_c(T)\, \epsilon^2\Bigr),
    \end{equation*}
which implies
    \begin{equation*}
        \epsilon \approx \left( \Delta_{\text{min}} - 2 c(T) \right) \sqrt{\frac{n(T)}{8N_c(T)}}.
    \end{equation*}

\noindent By the algorithm’s design, once $\alpha$ holds at time $T$ (so all confidence intervals are disjoint), any further conflict feedback becomes deterministic. \emph{As long as players keep attempting arms,} their empirical win estimates will converge to the true values. The following lemma establishes this. 

\begin{lemma}
    \label{lemma:deterministic_convergence}
    Let $T \ge 1$ be a positive integer and let $\hat{x}_T\in [0,1]$ be a real number, which we interpret as the average of $T$ samples. Let $p\in\{0,1\}$ be a fixed value (the “true” value). For any integer $t\geq 0$, define the updated average after adding $t$ additional samples (each equal to $p$) by
        \begin{equation*}
        \hat{x}_{T+t} \;=\; \frac{T\,\hat{x}_T + t\, p}{T+t}\,.
        \end{equation*}
    Then, the error after $t$ additional samples satisfies
        \begin{equation*}
        \left|\hat{x}_{T+t}-p\right| \;\leq\; \frac{T}{T+t}
        \end{equation*}
\end{lemma}

\begin{proof}
Starting from the definition of $\hat{x}(T+t)$, we have

\begin{align*}
    \hat{x}_{T+t} - p &= \frac{T\,\hat{x}_T + t\, p}{T+t} - p\\
                 &= \frac{T\,\hat{x}_T + t\, p - p(T+t)}{T+t}\\
                 &= \frac{T\,\hat{x}_T + t\, p - pT - pt}{T+t}\\
                 &= \frac{T\,(\hat{x}_T-p)}{T+t}
\end{align*}

Taking absolute values of both sides gives
\begin{align*}
    \left|\hat{x}(T+t)-p\right| &= \frac{T}{T+t}\,\left|\hat{x}-p\right|\\
                                & \le \frac{T}{T+t}
\end{align*}
\end{proof}

But realize that not all conflict win estimates will converge since some triplets eventually stop receiving feedback. In the context of Theorem \ref{thm:2}, before $\alpha(T)$, players are getting sample rewards as they are getting picked by the arms uniformly at random, leading to the players having some \emph{optimistic} reward estimate. Consequently, after $\alpha(T)$, players have some reward estimates for the arms and a belief about conflict wins (with the worst-case estimate being around $0.5 - \epsilon$). Right after $\alpha(T)$ occurs, the combined effect of the optimistic reward estimates and optimism function ensures that conflicts persist i.e. the players continue trying the arms. As more conflicts occur, the empirical win estimates gradually converge toward their true values, causing the impact of the optimism function to either be magnified (if the win probability exceeds 0.5, effectively guaranteeing a win) or diminished. This process enables players to gather more accurate information for selecting arms. In cases where a player already receives a better reward from one arm, they may cease exploring some inferior arm. This would cause the conflict win probability of the associated inferior triplet to never reach the true value. But as we will see in the following lemma, this is the case only for \emph{inconsequential} triplets. 

\begin{lemma}
    \label{lemma:inconsistency_resolution}
    Suppose the $\alpha(T)$ condition is achieved. Then an inconsistent triplet $(p_i, p_j, a_k)$ at time $T$ is either resolved, or is inconsequential by time $(T+n)$.
\end{lemma}

\begin{proof}
    Recall that an inconsistent triplet $(p_i, p_j, a_k)$ at time $T$ means one of the following holds for some $\epsilon > 0$:
    \begin{enumerate}
        \item $P_{ij}^k = 1$ but $\hat{P}_{ij}^k \le 1 - \epsilon$; 
        \item $P_{ij}^k = 0$ but $\hat{P}_{ij}^k \ge \epsilon$.
    \end{enumerate}
    In other words, player $p_i$'s empirical estimate $\hat{P}_{ij}^k$ is ``far'' from the true probability $P_{ij}^k \in \{0,1\}$. 
    Because the event $\alpha(T)$ has occurred, \textbf{all subsequent feedback from time $t$ onward is deterministic and aligns with the true outcome}. Since we condition on $\alpha(T)$, we preclude $p_i$ from \textbf{not attempting} $a_k$. We split into two cases, depending on whether the triplet $(p_i, p_j, a_k)$ remains relevant to $p_i$'s decisions:

    \paragraph{Case 1: The triplet remains relevant.} 
    If player $p_i$ continues to attempt arm $a_k$ and thus periodically conflicts with player $p_j$, then each time this conflict occurs, the empirical estimate $\hat{P}_{ij}^k$ receives noise-free feedback. Consequently:
    \begin{itemize}
        \item If $P_{ij}^k = 1$, $p_i$ always wins these conflicts. Hence, $\hat{P}_{ij}^k$ will rapidly increase above $1-\epsilon$, thus eliminating the inconsistency.
        \item If $P_{ij}^k = 0$, $p_i$ always loses. Hence, $\hat{P}_{ij}^k$ will drop below $\epsilon$, again removing the inconsistency.
    \end{itemize}
    By Lemma \ref{lemma:deterministic_convergence}, only a bounded number of outcomes is required before $\hat{P}_{ij}^k$ becomes accurate enough that $(p_i, p_j, a_k)$ is no longer inconsistent.

    \paragraph{Case 2: The triplet becomes inconsequential.}
    If instead player $p_i$ no longer selects arm $a_k$---for instance, because it discovers another arm $\smash{a_{k'}}$ with a higher UCB estimate or learns that $a_k$ is not profitable---then the conflict $(p_i, p_j, a_k)$ does not occur in future rounds. In that case, the belief about $P_{ij}^k$ no longer influences $p_i$'s \emph{decisions} and hence no longer impacts $p_i$'s rewards. We then say that $(p_i, p_j, a_k)$ is \emph{inconsequential}, because its incorrect empirical probability does not affect the algorithm's action selection.

    In either scenario, by at most $n$ additional time steps, any given inconsistent triplet is either corrected by deterministic feedback (Case 1) or becomes irrelevant to the player (Case 2). This shows that \emph{every} inconsistent triplet at time $t$ is resolved or inconsequential by time $t+n$.
\end{proof}

Now, we are ready to complete the proof of Theorem \ref{thm:2}. By design, once $\alpha$ occurs at time $T$, the feedback provided to the players is deterministic. Moreover, by Lemma \ref{lemma:optimism_all}, the probability that the optimism function lets the players continue attempting the arms despite bad conflict win estimates is lower bounded by: 
\begin{equation*}
1 - \underbrace{\frac{KN^2}{2} \exp\!\Bigl(-2\, N_c(T)\, \epsilon^2\Bigr)}_{\tau(t)},
\end{equation*}
where $\epsilon = \left( \Delta_{\text{min}} - 2 c(T) \right) \sqrt{\frac{n(T)}{8N_c(T)}}$ is chosen to match the tail probability in Theorem \ref{thm:1}. Finally, by applying Lemma \ref{lemma:deterministic_convergence}, we conclude that for all consequential conflicts $(p_i,p_j, p_k)$ and any $t \geq 0$, the updated empirical probability satisfies
\begin{equation*}
\left|\hat{P}_{ij}^k(T+t) - p_{ij}^k\right| \leq \frac{T}{T+t}.
\end{equation*}
\noindent And by Lemma \ref{lemma:inconsistency_resolution}, conflict estimates that fail to satisfy this inequality are inconsequential. This completes the proof of Theorem \ref{thm:2}. \qed

\noindent \textbf{Important: }Notice that as $t \to \infty, T \to \infty \implies n(T) \to \infty$. Because we condition on $\alpha(T)$, $N_c(T)$ is a constant. This implies, for a similar reason to $\delta(t) \to 0$, as $t \to \infty, \tau(t) \to 0$. This will later be important when establishing that PCA-SCA's algorithm will eventually be identical to CA-UCB as $t\to \infty$

\subsection{Proof of Theorem \ref{thm:3}}

The proof follows by combining Theorem~\ref{thm:1}, Theorem~\ref{thm:2}, and the supporting lemmas to show that, as $t\to\infty$, the execution of PCA-SCA converges to that of CA-UCB.

Theorem~\ref{thm:1} provides a probabilistic bound on the event $\alpha(T)^c$ (i.e. when the arms’ confidence intervals overlap and the win feedback remains non-deterministic). Theorem~\ref{thm:2} further bounds the event that the optimism function fails to motivate continued sampling, ensuring that every consequential triplet eventually receives sufficient feedback. Formally, define the failure event as things ``going wrong" as:

\begin{align*}
    P(\text{ things go wrong }) &= P( \text{ arms' confidence intervals ovelap } \cup \\
     & \text{ some consequential triplet has a win estimate less than } 0.5 - \epsilon) \\
     & \leq P( \text{ arms do not learn preferences }) +   \\
     & P( \text{ some triplet has a win estimate less than } 0.5 - \epsilon)\\
     & \leq \delta(t) + \tau(t) \left[\because \text{ Expressions from Theorem \ref{thm:1} and \ref{thm:2}} \right]
\end{align*}

\noindent where both $\delta(t)$ and $\tau(t)$ tend to zero as $t\to\infty$. Consequently, $P(\text{failure})\to 0$. Moreover, by Lemma~\ref{lemma:inconsistency_resolution}, for every consequential triplet the empirical conflict win probability converges arbitrarily close to its true value as $t\to\infty$. This implies that, in the limit, arms' have knowledge of their own preferences, and players have knowledge of arms' preferences (in terms of the win conflict probabilities). If we consider all the arms where the probability of winning a conflict is $ \geq 1 - \epsilon$ as the arms that go in the player's plausible set, players always choose the arm with the highest UCB estimate. It then follows that, with probability tending to one, the execution of PCA-SCA becomes identical to that of CA-UCB. 
\qed

\section{Alternative Methods for Controlling Player Preference Heterogeneity}
\label{appendix:Alternate_hetero}

There are alternative methods for controlling heterogeneity in player preferences, for example notions like uncoordinated/coordinated markets \cite{roglin_and_colleagues}. First, note that uncoordinated markets as defined in \cite{roglin_and_colleagues} essentially have uncorrelated preferences (equivalent to our model with low $\beta$), and coordinated markets introduce correlation, albeit using a different model than ours (with correlation defined on the edges of the matching graph rather than the vertices). 

\begin{figure}[H]
    \centering
    \includegraphics{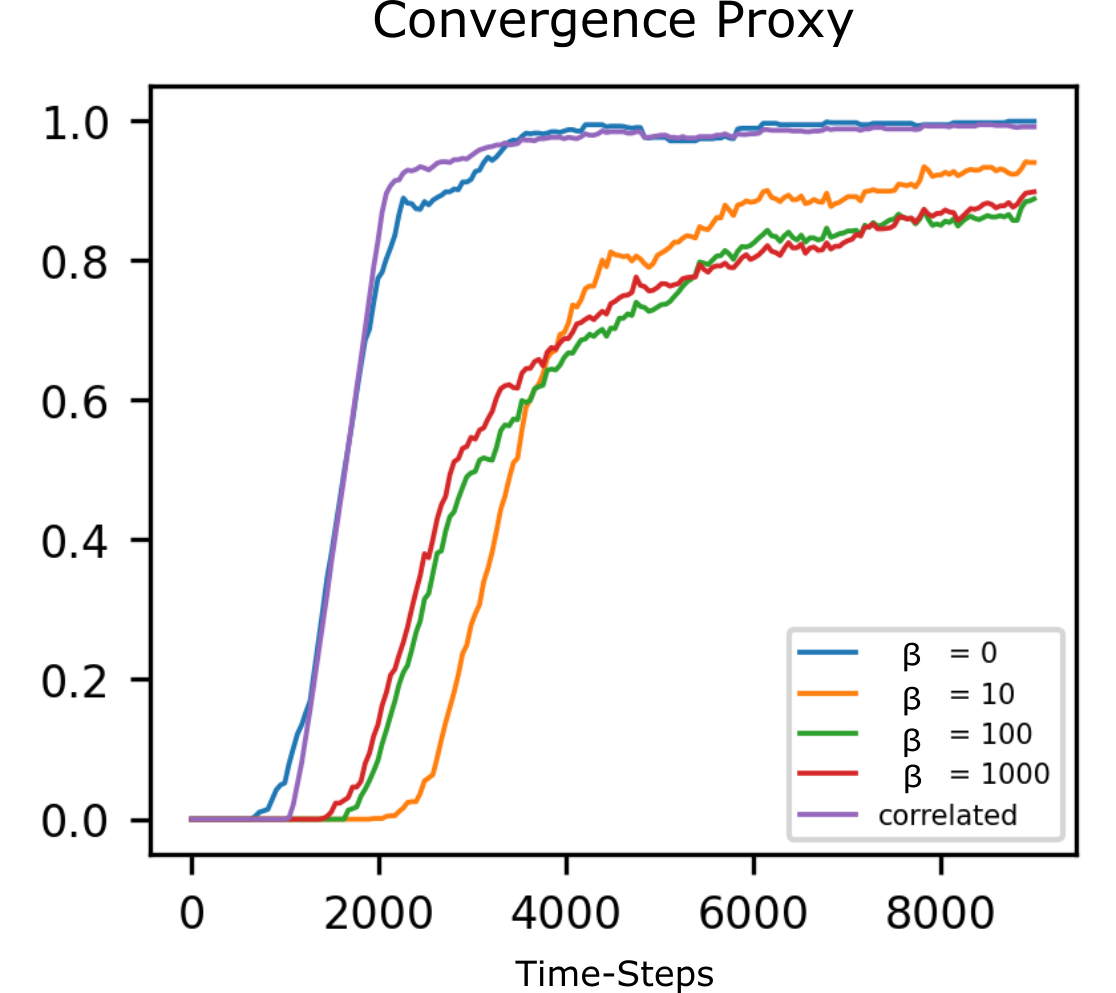}
    \caption{\begin{small}Convergence proxy for players with correlated preferences running OCA-UCB in the APKP model with $\mathcal{X} = 50, \theta = 90$ . The line in purple is the the edge-correlation as defined in \cite{roglin_and_colleagues} and the rest correspond to our paper. We can see that the  edge-correlated cases are comparable to when preferences are uniformly random (compare with node-correlation, where convergence time increases)\end{small}}
    \label{fig:appendixB}
\end{figure}

We ran experiments (with results averaged over 100 different runs for N=K=10) with edge-correlated preferences as defined in \cite{roglin_and_colleagues}, the results of which are shown in Figure \ref{fig:appendixB}.

Our findings show that the convergence time in these edge-correlated cases is actually comparable to when preferences are uniformly random (compare with node-correlation, where convergence time increases). 

\section{Visualization of the Optimism Function}
\label{appendix:optimism_viz}
\begin{figure}[h]
    \centering
    \includegraphics[width=0.4\linewidth]{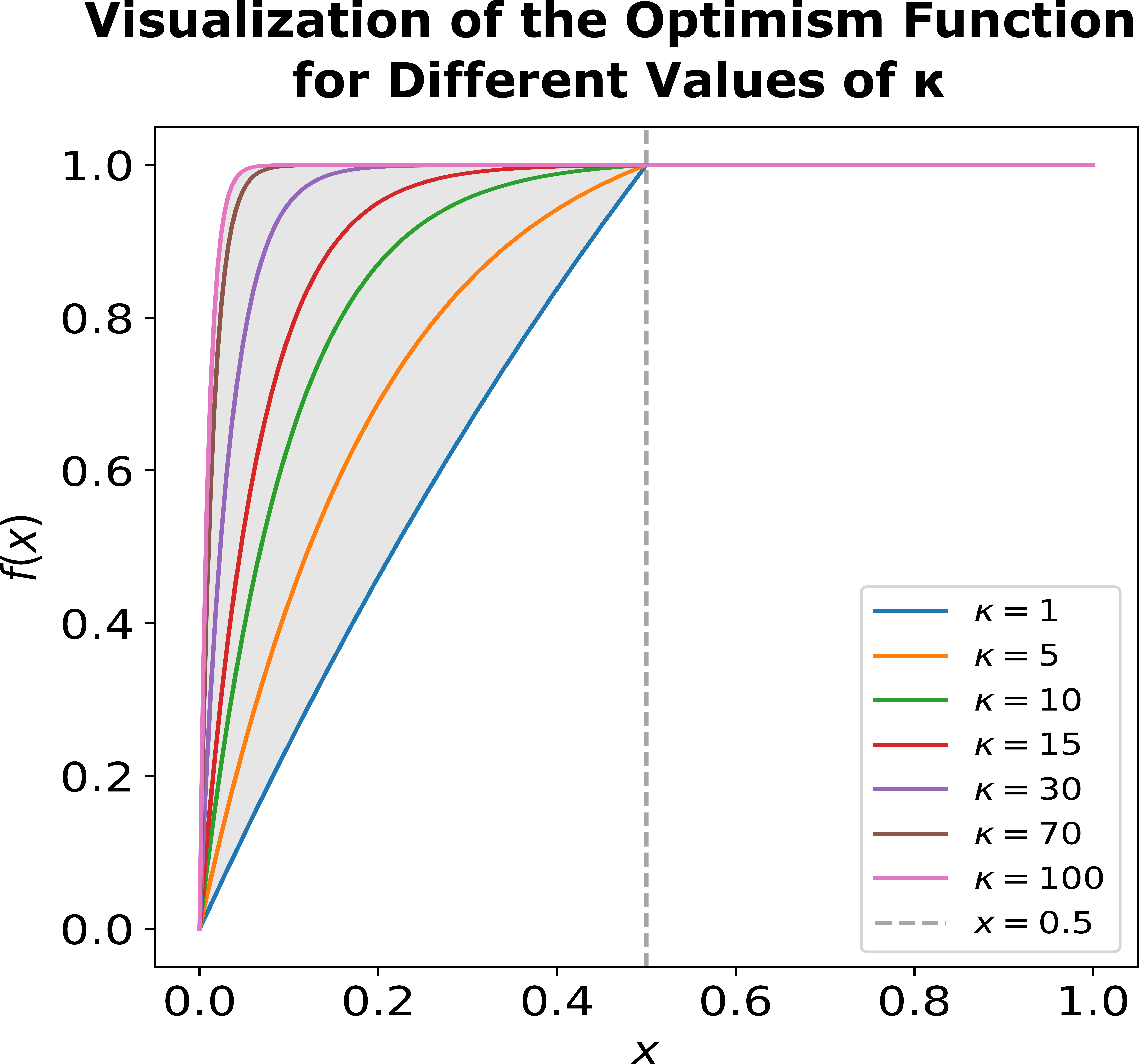}
    \caption{Visualization of different levels of ``optimism'' for different values of $\kappa$. Note that as $\kappa$ increases, the players are more optimistic about their chances of winning a conflict if if their empirical estimate is below 0.5}
    \label{fig:optimism}
\end{figure}
\section{Machine Specifications}
\label{appendix:specs}

All experiments were run on a \textbf{Apple MacBook Pro (16-inch, 2021) } with the M1 Pro chip. Technical specifications:

\begin{itemize}
  \item \textbf{Year introduced:} 2021
  \item \textbf{Chip:}
    \begin{itemize}
      \item Apple M1 Pro: 10-core CPU (8 performance + 2 efficiency cores)
      \item 16-core GPU; 16-core Neural Engine
      \item 200 GB/s unified memory bandwidth
      \item Media engine: HW-accelerated H.264, HEVC, ProRes \& ProRes RAW; video decode/encode; ProRes encode/decode
    \end{itemize}
  \item \textbf{Memory:} 16 GB unified (on-chip) RAM
  \item \textbf{Storage:} 512 GB SSD
  \item \textbf{Battery \& Power:}
    \begin{itemize}
      \item 100 Wh lithium-polymer battery
    \end{itemize}
\end{itemize}

\section{Statistical Analysis of Time to Convergence}
\label{appendix:significance_testing}

To generate the Graphs in Figure 3, setting $\mathcal{X} = 1000$ and $\theta = 90$, and calculate the convergence proxy. Here, we aim to test statistically if PCA-TS converges faster than PCA-TS.We proceed as follows.  For each run $j$ and market size $N$, define
    $$
        T^{(j)} = \min\bigl\{\,t : \text{convergence proxy}_j(t) = 1\bigr\},
    $$
and let the paired difference be:
    $$
        \Delta^{(j)} \;=\; T_{\mathrm{UCB}}^{(j)} \;-\; T_{\mathrm{TS}}^{(j)}.
    $$
Because the distribution of $\Delta^{(j)}$ is right–skewed and contains imputed maximum values for runs that never converged within the horizon, we use two nonparametric tests:
\begin{enumerate}
  \item \textbf{Wilcoxon signed‐rank test} for 
    $
      H_0:\ \mathrm{median}(\Delta)=0
      \quad\text{vs.}\quad
      H_1:\ \mathrm{median}(\Delta)>0,
    $
    which assesses whether UCB takes longer than TS in median.
  \item \textbf{Binomial sign test} for 
    $
      H_0:\ \Pr(T_{\mathrm{TS}} < T_{\mathrm{UCB}}) = 0.5
      \quad\text{vs.}\quad
      H_1:\ \Pr(T_{\mathrm{TS}} < T_{\mathrm{UCB}}) > 0.5,
    $
    which checks if TS is faster in a majority of runs.
\end{enumerate}

Table~\ref{tab:convergence_results} reports, for each market size $N$, the median difference $\mathrm{median}(\Delta)$, the Hodges–Lehmann estimator of the shift, and the $p$-values from both tests.

\begin{table}[ht]
\centering
\caption{Time-to-convergence summary by market size \(N\).}
\label{tab:convergence_results}
\begin{tabular}{cccc}
\toprule
\(N\) & \(\mathrm{median}(\Delta)\) & Wilcoxon \(p\) & Binomial \(p\) \\
\midrule
5  & 577 & \(5.7\times10^{-10}\) & \(3.1\times10^{-11}\) \\
10 & 890 & \(4.6\times10^{-9}\)  & \(2.8\times10^{-8}\)  \\
15 & 561 & \(9.2\times10^{-6}\)  & \(1.6\times10^{-5}\)  \\
20 & 816 & \(7.7\times10^{-4}\)  & \(2.3\times10^{-3}\)  \\
\bottomrule
\end{tabular}
\end{table}

\noindent
\textbf{Results.}  For all tested market sizes \(N\), Thompson Sampling reaches the convergence proxy of 1 significantly faster than PCA-UCB under both the Wilcoxon signed-rank test and the binomial sign test (all $p<0.01$ ). Median speed-ups range from 561 to 890 time steps.


\end{document}